\documentclass[graybox]{svmult}

\usepackage{helvet}         
\usepackage{courier}        
\usepackage{type1cm}        
\usepackage{makeidx}         
\usepackage{graphicx}        
\usepackage{multicol}        
\usepackage[bottom]{footmisc}

\makeindex             

\usepackage{bm}
\newcommand{\square}{\kern1pt\vbox{\hrule height
1.2pt\hbox{\vrule width 1.2pt\hskip 3pt
   \vbox{\vskip 6pt}\hskip 3pt\vrule width 0.6pt}\hrule
height 0.6pt}\kern1pt}
\def\Mpl{M_{\rm pl}}

\newcommand{\C}{{\cal C}}
\newcommand{\s}{{\cal S}}
\newcommand{\Z}{{\cal Z}}
\newcommand{\U}{{\cal U}}
\newcommand{\A}{{\cal A}}
\newcommand{\B}{{\cal B}}
\newcommand{\G}{{\cal G}}
\newcommand{\M}{{\cal M}}

\begin{document}

\newcommand{\vp}{\varphi}
\newcommand{\rd}{{\rm d}}
\newcommand{\gsim}{\mbox{\raisebox{-1.ex}{$\stackrel
     {\textstyle>}{\textstyle\sim}$}}}
\newcommand{\lsim}{\mbox{\raisebox{-1.ex}{$\stackrel
     {\textstyle<}{\textstyle \sim}$}}}

\title*{{\bf The effective field theory of inflation/dark energy and 
the Horndeski theory}}

\author{Shinji Tsujikawa}

\institute{Department of Physics, 
Faculty of Science,
Tokyo University of Science,
1-3 Kagurazaka, Shinjuku-ku, 
Tokyo, 162-8601, Japan \email{shinji@rs.kagu.tus.ac.jp}}

\maketitle

\abstract{
The effective field theory (EFT) of cosmological perturbations is a useful 
framework to deal with the low-energy degrees of freedom present for 
inflation and dark energy. We review the EFT for modified gravitational 
theories by starting from the most general action in unitary gauge 
that involves the lapse function and  the three-dimensional geometric scalar quantities 
appearing in the Arnowitt-Deser-Misner (ADM) formalism. 
Expanding the action up to quadratic order in the 
perturbations and imposing conditions for the elimination of 
spatial derivatives higher than second order, we obtain the Lagrangian 
of curvature perturbations and gravitational waves with a single scalar degree 
of freedom. The resulting second-order Lagrangian is exploited
for computing the scalar and tensor power spectra generated 
during inflation. We also show that the most general 
scalar-tensor theory with second-order equations of 
motion--Horndeski theory--belongs to the action of our general 
EFT framework and that the background equations 
of motion in Horndeski theory can be conveniently expressed
in terms of three EFT parameters. Finally we study the equations of 
matter density perturbations and the effective gravitational coupling 
for dark energy models based on Horndeski theory, to confront 
the models with the observations of large-scale structures 
and weak lensing. 
}


\section{Introduction}	

The inflationary paradigm, which was originally proposed to 
solve a number of cosmological problems in the standard 
Big Bang cosmology \cite{Sta80,oldinf}, 
is now widely accepted as a viable 
phenomenological framework describing the accelerated 
expansion in the early Universe. In particular, the 
Cosmic Microwave Background (CMB) temperature 
anisotropies measured by COBE \cite{COBE}, 
WMAP \cite{WMAP1}, and Planck \cite{Planck} 
satellites support the slow-roll inflationary scenario
driven by a single scalar degree of freedom. 
Inflation generally predicts the nearly scale-invariant 
primordial power spectrum of curvature perturbations \cite{oldper}, 
whose property is consistent with the observed 
CMB anisotropies. In spite of its great success, 
we do not yet know the origin of the scalar field 
responsible for inflation (dubbed ``inflaton'').

The observations of the type Ia 
Supernovae (SN Ia) \cite{Riess,Perlmutter} showed that 
the Universe entered the phase of another accelerated 
expansion after the matter-dominated epoch. 
This has been also supported by other independent 
observations such as CMB \cite{WMAP1} and 
Baryon Acoustic Oscillations (BAO) \cite{BAO}. 
The origin of the late-time cosmic acceleration 
(dubbed ``dark energy'') is not identified yet.
The simplest candidate for dark energy is the cosmological constant 
$\Lambda$, but if it originates from the vacuum energy appearing 
in particle physics, the theoretical value is enormously larger 
than the observed dark energy scale \cite{Weinberg,CST}. 
There is a possibility that some scalar degree of freedom 
(like inflaton) is responsible for dark energy \cite{quinpapers}. 

Although many models of inflation and dark energy have been 
constructed in the framework of General Relativity (GR), 
the modification of gravity from GR can also give rise 
to the epoch of cosmic acceleration. 
For example, the Starobinsky model 
characterized by the Lagrangian $f(R)=R+R^2/(6M^2)$ \cite{Sta80}, 
where $R$ is a Ricci scalar and $M$ is a constant, leads to the
quasi de Sitter expansion of the Universe.
The recent observational constraints on the dark energy equation 
of state $w_{\rm DE}=P_{\rm DE}/\rho_{\rm DE}$ (where 
$P_{\rm DE}$ and $\rho_{\rm DE}$ is the pressure and 
the energy density of dark energy respectively) imply that the region 
$w_{\rm DE}<-1$ is favored from the joint data analysis 
of SN Ia, CMB, and BAO \cite{WMAP9,CDT,Planck}. 
If we modify gravity from GR, it is possible to realize 
$w_{\rm DE}<-1$ without having a problematic 
ghost state (see Refs.~\cite{moreview} for reviews). 

Given that the origins of inflation and dark energy have not been 
identified yet, it is convenient to construct a general 
framework dealing with gravitational degrees 
of freedom beyond GR.
In fact, the EFT of inflation and dark energy 
provides a systematic parametrization that  
accommodates possible low-energy degrees of 
freedom by employing cosmological 
perturbations as small expansion parameters about the 
Friedmann-Lema\^{i}tre-Robertson-Walker (FLRW) 
background \cite{Cremi,Cheung,Quin}. 
This EFT approach allows one to facilitate the confrontation 
of models with the cosmological data.

Originally, the EFT of inflation was developed to quantify 
high-energy corrections to the standard slow-roll inflationary 
scenario \cite{Weinberg2}.
Expanding the action up to third order in the cosmological 
perturbations, it is also possible to estimate higher-order correlation 
functions associated with primordial non-Gaussianities \cite{ng1}. 
The EFT formalism was applied to dark energy in connection 
to the large-distance modification of gravity \cite{Park}-\cite{Gergely}.
The advantage of this approach is that practically all the 
single-field models of inflation and dark energy can be 
accommodated in a unified way \cite{Bloom}.

Starting from the most general action that depends on the lapse 
function and other geometric three-dimensional scalar quantities present 
in the ADM formalism, Gleyzes {\it et al.} \cite{Piazza}
expanded the action up to quadratic order in cosmological perturbations
of the ADM variables. In doing so, the perturbation $\delta \phi$ of a 
scalar field $\phi$ can be generally present, but the choice of 
unitary gauge ($\delta \phi=0$) allows one to absorb the field 
perturbation in the gravitational sector.
Once we fix the gauge in this way, introducing another 
scalar-field perturbation implies that the system possesses 
at least two-scalar degrees of freedom. 
In fact, such a multi-field scenario was studied in 
Ref.~\cite{Gergely} to describe both dark energy 
and dark matter.

By construction, the EFT formalism developed in 
Refs.~\cite{Cremi,Cheung,Bloom2,Piazza} 
keeps the time derivatives under control, while the spatial derivatives higher 
than second order are generally present. Imposing conditions to 
eliminate these higher-order spatial derivatives for the general theory 
mentioned above, Gleyzes {\it et al.} \cite{Piazza} derived the quadratic 
Lagrangian of cosmological perturbations with one scalar 
degree of freedom. If the scalar degree of freedom is responsible 
for inflation, for example, the resulting power spectrum 
of curvature perturbations can be computed on the quasi 
de Sitter background (along the same lines in 
Refs.~\cite{Muka,KYY,Tsujinon,XGao,Tsujinon2}).
In this review, we evaluate the inflationary power spectra of both 
scalar and tensor perturbations expressed in terms of 
the ADM variables.

In 1973, Horndeski derived the action of the most general
scalar-tensor theories with second-order equations of motion \cite{Horndeski}. 
This theory recently received much attention as an extension of 
(covariant) Galileons \cite{Nicolis,cova,Char}. 
One can show that the four-dimensional action of
``generalized Galileons'' derived by Deffayet {\it et al.} \cite{Deffayet} 
is equivalent to the Horndeski 
action after a suitable field redefinition \cite{KYY}. 
Gleyzes {\it et al.} \cite{Piazza} expressed the Horndeski Lagrangian 
in terms of the ADM variables appearing 
in the EFT formalism (see also Ref.~\cite{Bloom2}).
This allows one to have a connection between the
Horndeski theory and the EFF of inflation/dark energy. 
In fact, it was shown that Horndeski theory belongs to a sub-class 
of the general EFT action \cite{Piazza}.

For the background cosmology, the EFT of inflation/dark energy 
is characterized by three time-dependent parameters 
$f$, $\Lambda$, and $c$ \cite{Cremi,Cheung,Quin}. 
This property is useful to perform general analysis 
for the dynamics of dark energy \cite{Silve}. 
In the EFT of dark energy, Gleyzes {\it et al.} \cite{Piazza} 
obtained the equations of linear cosmological perturbations 
in the presence of non-relativistic matter (dark matter, baryons). 
This result reproduces the perturbation equations in Horndeski theory
previously derived in Ref.~\cite{Koba}.
We note that the perturbation equations in the presence of 
another scalar field (playing the role of dark matter) were 
also derived in Ref.~\cite{Gergely}. 
These results are useful to confront modified gravitational models 
of dark energy with the observations of large-scale structures, 
weak lensing, and CMB.

In this lecture note, we review the EFT of inflation/dark energy 
following the recent works of Refs.~\cite{Piazza,Gergely}.  

In Sec.~\ref{actionsec} we start from a general 
gravitational action in unitary gauge and derive the background 
equations of motion on the flat FLRW background.

In Sec.~\ref{secondsec} we obtain the linear perturbation 
equations of motion and discuss conditions for avoiding 
ghosts and Laplacian instabilities of scalar and tensor perturbations.

In Sec.~\ref{infsec} the inflationary power spectra of scalar 
and tensor perturbations are derived for general 
single-field theories with second-order linear perturbation
equations of motion.

In Sec.~\ref{Hornsec} we introduce the action of Horndeski theory 
and express it in terms of the ADM variables appearing in the 
EFT formalism.

In Sec.~\ref{effsec} we discuss how the second-order EFT action accommodates
Horndeski theory as specific cases and provide the correspondence 
between them.

In Sec.~\ref{darksec} we apply the EFT formalism to dark energy 
and obtain the background equations of motion in a generic way.
In Horndeski theory, the equations of matter density perturbations and 
the effective gravitational coupling are derived in the presence 
of non-relativistic matter.

Sec.~\ref{consec} is devoted to conclusions.

Throughout the paper we use units such that $c=\hbar=1$,
where $c$ is the speed of light and $\hbar$
is reduced Planck constant. 
The gravitational constant $G$ is related to 
the reduced Planck mass
$M_{{\rm pl}}=2.4357\times10^{18}$\,GeV via 
$8\pi G=1/M_{{\rm pl}}^{2}$. 
The Greek and Latin indices represent components in 
space-time and in a three-dimensional 
space-adapted basis, respectively. 
For the covariant derivative of some physical quantity $Y$, 
we use the notation $Y_{;\mu}$ or $\nabla_{\mu} Y$.
We adopt the metric signature $(-,+,+,+)$.

\section{The general gravitational action in unitary gauge and 
the background equations of motion}	
\label{actionsec}

The EFT of cosmological perturbations allows one to deal with the 
low-energy degree of freedom appearing for inflation and dark energy.
In particular, we are interested in the minimal extension of GR 
to modified gravitational theories with a single scalar degree 
of freedom $\phi$.
The EFT approach is based on the choice of unitary gauge in which 
the constant time hypersurface coincides with the constant $\phi$ 
hypersurface. In other words, this corresponds to the gauge choice
\begin{equation}
\delta \phi=0\,,
\label{unitary}
\end{equation}
where $\delta \phi$ is the field perturbation. 
In this gauge the dynamics of $\delta \phi$ is ``eaten'' by the metric, 
so the Lagrangian does not have explicit $\phi$ dependence 
about the flat FLRW background.

The EFT of cosmological perturbations is based on the $3+1$ decomposition 
of the ADM formalism \cite{ADM}. In particular, the $3+1$ splitting in unitary gauge 
allows one to keep the number of time derivatives under control, while 
higher spatial derivatives can be generally present. 
As we will see later, this property is especially useful for constructing 
theories with second-order time and spatial derivatives.
The ADM line element is given by
\begin{equation}
ds^{2}=g_{\mu \nu }dx^{\mu }dx^{\nu}
=-N^{2}dt^{2}+h_{ij}(dx^{i}+N^{i}dt)(dx^{j}+N^{j}dt)\,,  
\label{ADMmetric}
\end{equation}
where $N$ is the lapse function, $N^i$ is the shift vector, and
$h_{ij}$ is the three-dimensional metric.
Then,  the four-dimensional metric $g_{\mu \nu}$ can be expressed 
as $g_{00}=-N^2+N^iN_i$, $g_{0i}=g_{i0}=N_i$, and $g_{ij}=h_{ij}$. 
A unit vector orthogonal to the constant $t$ hypersurface $\Sigma_t$ 
is given by $n_{\mu}=-Nt_{;\mu}=(-N,0,0,0)$, and
hence $n^{\mu}=(1/N,-N^{i}/N)$ with $n_{\mu}n^{\mu}=-1$.
The induced metric $h_{\mu \nu}$ on $\Sigma_t$ can be 
expressed as $h_{\mu \nu}=g_{\mu \nu}+n_{\mu}n_{\nu}$, 
so that it satisfies the orthogonal relation 
$n^{\mu} h_{\mu \nu}=0$.

The extrinsic curvature is defined by 
\begin{equation}
K_{\mu \nu}=h^{\lambda}_{\mu} n_{\nu;\lambda}
=n_{\nu;\mu}+n_{\mu}a_{\nu}\,,
\label{Kdef}
\end{equation}
where $a^{\nu}=n^{\lambda}n^{\nu}_{;\lambda}$ is the acceleration 
(curvature) of the normal congruence $n^{\nu}$.
Since there is the relation $n^{\mu}K_{\mu \nu}=0$, 
the extrinsic curvature is the quantity on $\Sigma_t$. 
The internal geometry of $\Sigma_t$ can be quantified 
by the three-dimensional Ricci tensor 
${\cal R}_{\mu \nu} \equiv {}^{(3)}R_{\mu \nu}$ 
associated with the metric $h_{\mu \nu}$.
The three-dimensional Ricci scalar 
${\cal R}={{\cal R}^{\mu}}_{\mu}$ is related to 
the four-dimensional Ricci scalar $R$, as
\begin{equation}
R={\cal R}+K_{\mu \nu}K^{\mu \nu}-K^2
+2 ( Kn^{\mu}-a^{\mu} )_{;\mu}\,,
\label{Rre}
\end{equation}
where $K \equiv {K^{\mu}}_{\mu}$ is the trace of 
the extrinsic curvature.

In the following we study general gravitational theories 
that depend on scalar quantities appearing
in the ADM formalism \cite{Bloom,Bloom2,Piazza}. 
In addition to the lapse $N$, we have the following scalars 
\begin{equation}
K\equiv {K^{\mu}}_{\mu}\,,\quad 
\s \equiv K_{\mu \nu }K^{\mu \nu}\,,\quad 
\mathcal{R}\equiv {\mathcal{R}^{\mu}}_{\mu}\,,\quad 
\Z \equiv \mathcal{R}_{\mu \nu}\mathcal{R}^{\mu \nu}\,,\quad 
\U \equiv \mathcal{R}_{\mu \nu}K^{\mu \nu}\,.  \label{threedef}
\end{equation}
The Lagrangian $L$ of general gravitational theories 
depends on these scalars, so that the action is given by 
\begin{equation}
S=\int d^{4}x\sqrt{-g}\,L(N,K,\s,{\cal R},
\Z,\U;t)\,.\label{action0}
\end{equation}
We do not include the dependence of the scalar quantity 
${\cal N}=N^{\mu} N_{\mu}$ coming from the shift vector, 
since such a term does not appear even in the most general 
scalar-tensor theories with second-order equations of motion 
(see Sec.~\ref{Hornsec}). 
In the action (\ref{action0}), the time dependence is also explicitly 
included because in unitary gauge its dependence is directly 
related to the scalar degree of freedom, such that $\phi=\phi(t)$.
The field kinetic term\footnote{We caution that the notation of 
the field kinetic energy is the same as that used in 
Refs.~\cite{Piazza,Gergely}, but the notation of $X$ 
used in Refs.~\cite{KYY,Tsujinon,XGao,Tsujinon2,Koba} is $-1/2$ times different.} 
\begin{equation}
X \equiv g^{\mu \nu}\partial_{\mu}\phi 
\partial_{\nu} \phi
\end{equation}
depends on the lapse $N$ and the time $t$.
The field $\phi$ enters the equations of motion through the 
partial derivatives $L_{N} \equiv \partial L/\partial N$ and $
L_{NN} \equiv \partial ^{2}L/\partial N^{2}$.

Let us consider four scalar metric perturbations $A$, 
$\psi $, $\zeta $, and $E$ about the flat FLRW background 
with the scale factor $a(t)$. 
The general perturbed metric is given by 
\begin{equation}
ds^{2}=-e^{2A}dt^{2}+2\psi _{|i}dx^{i}dt+a^{2}(t) 
(e^{2\zeta }\delta_{ij}+\partial_{ij}E ) dx^{i}dx^{j}\,,
\label{permet}
\end{equation}
where $|i$ represents a covariant derivative with respect 
to $h_{ij}$, and $\partial_{ij} \equiv \nabla_i \nabla_j-\delta_{ij}\nabla^2/3$.
Under the transformation $t\rightarrow t+\delta t$ and $x^{i}\rightarrow
x^{i}+\delta ^{ij}\partial _{j}\delta x$, the perturbations $\delta \phi$ 
and $E$ transform as 
\begin{equation}
\delta \phi \rightarrow \delta \phi -\dot{\phi}\,\delta t\,,\qquad
E\rightarrow E-\delta x\,,
\end{equation}
where a dot represents a derivative with respect to $t$.
Choosing the unitary gauge (\ref{unitary}), 
the time slicing $\delta t$ is fixed. 
The spatial threading $\delta x$ can be fixed 
with the gauge choice
\begin{equation}
E=0\,.
\label{gauge2}
\end{equation}

On the flat FLRW background with the line element 
$ds^2=-dt^{2}+a^{2}(t)\delta _{ij}dx^{i}dx^{j}$, the 
three-dimensional geometric quantities are given by 
\begin{equation}
\bar{K}_{\mu \nu}=H\bar{h}_{\mu \nu}\,,
\quad \bar{K}=3H\,,\quad 
\bar{\s}=3H^{2}\,,\quad \bar{{\cal R}}_{\mu \nu}=0\,,
\quad \bar{{\cal R}}=\bar{\Z}=\bar{\U}=0\,,
\end{equation}
where a bar represents background values and 
$H \equiv \dot{a}/a$ is the Hubble parameter.
We define the following perturbed quantities
\begin{equation}
\delta K_{\nu}^{\mu}=K_{\nu }^{\mu}-Hh_{\nu }^{\mu }\,,
\quad \delta K=K-3H\,,\quad 
\delta \s=\s-3H^{2}=2H\delta K+\delta
K_{\nu }^{\mu} \delta K_{\mu}^{\nu}\,,
\label{delKS}
\end{equation}
where the last equation arises from the first equation and 
the definition of $\s$. Since $\cal{R}$ and $\Z$ vanish on the
background, they appear only as perturbations. 
Up to quadratic order in perturbations, 
they can be expressed as
\begin{equation}
\delta \mathcal{R}=\delta _{1}\mathcal{R}+\delta _{2}\mathcal{R}\,,
\qquad
\delta \Z=\delta \mathcal{R}_{\nu }^{\mu}\delta 
\mathcal{R}_{\mu}^{\nu}\,,
\end{equation}
where $\delta _{1} \cal{R}$ and $\delta _{2} \cal{R}$ are first-order
and second-order perturbations in $\delta \cal{R}$, respectively. 
The perturbation $\Z$ is higher than first order. 
The first equality (\ref{delKS}) implies 
\begin{equation}
\U=H{\cal R}+{\cal R}_{\nu}^{\mu}
\delta K_{\mu}^{\nu}\,,
\end{equation}
where the last term is a second-order quantity. 

In order to derive the background and perturbation 
equations of motion, we expand the action (\ref{action0}) 
up to quadratic order in perturbations, as
\begin{eqnarray}
\hspace{-0.5cm}
L &=&\bar{L}+L_{N}\delta N+L_{K}\delta K+L_{\s}
\delta \s+L_{\mathcal{R}}\delta \mathcal{R}
+L_{\Z}\delta \Z+L_{\U}
\delta \U \nonumber \\
\hspace{-0.5cm}
&&+\frac{1}{2}\left( \delta N\frac{\partial }{\partial N}
+\delta K\frac{\partial }{\partial K}+\delta \s
\frac{\partial }{\partial \s}+\delta \mathcal{R}
\frac{\partial }{\partial \mathcal{R}}+\delta \mathcal{Z}
\frac{\partial }{\partial \Z}+\delta \U
\frac{\partial }{\partial \U} \right)^2L\,,
\label{lag}
\end{eqnarray}
where a lower index of the Lagrangian $L$
denotes the partial derivatives with respect to the scalar quantities
represented in the index.
{}From the second and third relations of Eq.~(\ref{delKS}),
the expansion of the term $L_{K}\delta K+L_{\s}\delta \s$ 
up to second order reads
\begin{eqnarray}
L_{K}\delta K+L_{\s}\delta \s &=&{\cal F}(K-3H)
+L_{\s}\delta K_{\nu}^{\mu}\delta K_{\mu}^{\nu}  \nonumber \\
&\simeq& 
-\dot{{\cal F}}-3H\mathcal{F}+\dot{\mathcal{F}}\delta N
+L_{\s}\delta K_{\nu }^{\mu}\delta K_{\mu}^{\nu}
-\dot{\mathcal{F}}\delta
N^{2}\,,  \label{LKSre}
\end{eqnarray}
where 
\begin{equation}
\mathcal{F}\equiv L_{K}+2HL_{\s}\,.
\label{Fdef}
\end{equation}
In the second line of Eq.~(\ref{LKSre}), 
the term $\mathcal{F}K$ has been integrated 
by using the relation $K=n_{~;\mu }^{\mu}$, as
\begin{equation}
\int d^{4}x\sqrt{-g}\,\mathcal{F}K=-\int d^{4}x\sqrt{-g}\,n^{\mu }
\mathcal{F}_{;\mu}=-\int d^{4}x\sqrt{-g}\frac{\dot{\mathcal{F}}}{N}\,,
\label{delK}
\end{equation}
where the boundary term is dropped. 
Note that we have also expanded the term 
$N^{-1}=(1+\delta N)^{-1}$ up to second order 
in Eq.~(\ref{LKSre}).

The term $\U$ satisfies the relation 
\begin{equation}
\alpha(t) \U= \frac12 \alpha(t) {\cal R}K+
\frac{1}{2N} \dot{\alpha}(t) {\cal R}\,,
\end{equation}
where $\alpha(t)$ is an arbitrary function of $t$.
Using this relation and the fact that $\U$ is 
a perturbed quantity, it follows that  
\begin{eqnarray}
L_{\U} \delta \U
&=& \frac{1}{2}\left( \dot{L_{\U}}
+3HL_{\U}\right) \delta _{1}\mathcal{R}
+\frac{1}{2}\left( L_{\U}\delta K-\dot{L_{\U}}\delta N\right) \delta
_{1}\mathcal{R} \nonumber \\
& &+\frac{1}{2}\left( \dot{L_{\U}}
+3HL_{\U}\right) \delta _{2}\mathcal{R}\,,
\end{eqnarray}
where the first term on the r.h.s. is the first-order quantity, 
whereas the rest is second-order.

Summing up the terms discussed above, the zeroth-order and first-order 
Lagrangians of (\ref{lag}) are given, respectively, by 
\begin{eqnarray}
L_{0} &=&\bar{L}-\dot{\mathcal{F}}-3H\mathcal{F}\,,  \\
L_{1} &=&(\dot{\mathcal{F}}+L_{N})\delta N
+\mathcal{E}\delta _{1}\mathcal{R}\,, 
\label{lagfirst}
\end{eqnarray}
where 
\begin{equation}
\mathcal{E} \equiv L_{\cal R}+\frac12 \dot{L_\U}+
\frac32 H L_{\U}\,.
\end{equation}
Defining the Lagrangian density as $\mathcal{L}=\sqrt{-g}L=N\sqrt{h}\,L$,
where $h$ is the determinant of the three-dimensional metric $h_{ij}$, 
the zeroth-order and first-order terms read
\begin{eqnarray}
\mathcal{L}_{0} &=& a^3 \left( \bar{L}-\dot{\cal F}-3H {\cal F} \right)\,,
\label{lag0th} \\
\mathcal{L}_{1} &=&
a^{3}\left( \bar{L}+L_{N}-3H\mathcal{F}\right) \delta N
+\left( \bar{L}-\dot{\mathcal{F}}-3H\mathcal{F}\right)
\delta \sqrt{h}+
a^3\mathcal{E} \delta_1 {\cal R}\,. \label{L1}
\end{eqnarray}
The last term is a total derivative, so it can be dropped.
Varying the first-order Lagrangian (\ref{L1}) with respect to 
$\delta N$ and $\delta \sqrt{h}$, we can derive the following 
equations of motion respectively:
\begin{eqnarray}
&&\bar{L}+L_{N}-3H\mathcal{F}=0\,,
\label{back1} \\
&&\bar{L}-\dot{\mathcal{F}}-3H\mathcal{F}=0\,.
\label{back2} 
\end{eqnarray}
On using Eq.~(\ref{back2}), the zero-th order Lagrangian density 
(\ref{lag0th}) vanishes. 
Subtracting Eq.~(\ref{back1}) from Eq.~(\ref{back2}), we obtain
\begin{equation}
L_N+\dot{\cal F}=0\,.
\label{back3} 
\end{equation}
Two of Eqs.~(\ref{back1})-(\ref{back3}) determine the cosmological 
dynamics on the flat FLRW background.

As an example, let us consider the non-canonical scalar-field 
model given by \cite{kinf,kes}
\begin{equation}
L=\frac{M_{\mathrm{pl}}^{2}}{2}R+P(\phi ,X)\,,  
\label{klag}
\end{equation}
where $P$ is an arbitrary function with respect to $\phi$ and $X$. 
Using Eq.~(\ref{Rre}) and dropping the total divergence term, 
it follows that 
\begin{equation}
L=\frac{M_{\mathrm{pl}}^{2}}{2}
\left( \mathcal{R}+\s-K^{2}\right)
+P(\phi ,X)\,,
\label{multilag2}
\end{equation}
where $X=-N^{-2} \dot{\phi}^2$. 
Since $\bar{L}=-3M_{\mathrm{pl}}^{2}H^{2}+P$,
$L_{N}=2\dot{\phi}^{2}P_{X}$, and $\mathcal{F}=-2M_{\mathrm{pl}}^{2}H$ 
on the flat FLRW background, Eqs.~(\ref{back1}) and (\ref{back3}) read
\begin{eqnarray}
3M_{\mathrm{pl}}^{2}H^{2} &=&
-2P_{X}\dot{\phi}^{2}-P\,,
\label{back1d} \\
M_{\mathrm{pl}}^{2}\dot{H} &=&
\dot{\phi}^2P_{X}\,,
\label{back2d} 
\end{eqnarray}
which match with those derived in Ref.~\cite{kinf}.
Taking the time derivative of Eq.~(\ref{back1d}) and using 
Eq.~(\ref{back2d}), we obtain the field equation of motion 
\begin{equation}
\frac{d}{dt}\left( a^{3}P_{X}\dot{\phi}^{2}\right)
+\frac{1}{2}a^{3}\dot{P}=0\,,
\end{equation}
which is equivalent to 
$\frac{d}{dt}\left( a^{3}P_{X}\dot{\phi}\right) +\frac{1}{2}a^{3}P_{\phi
}=0$. For a canonical field characterized by the Lagrangian 
$P=-X/2-V(\phi)$, this reduces to the well-known equation 
$\ddot{\phi}+3H \dot{\phi}+V_{\phi}=0$.

\section{Second-order action for cosmological perturbations}	
\label{secondsec}

In order to derive the equations of motion for linear cosmological 
perturbations, we need to expand the action (\ref{action0}) 
up to quadratic order.
The Lagrangian (\ref{lag}) reads
\begin{eqnarray}
L &=&\bar{L}-\dot{\mathcal{F}}-3H\mathcal{F}
+(\dot{\mathcal{F}}+L_{N})\delta
N+\mathcal{E}\delta _{1}\mathcal{R} \nonumber \\
&&+\left( \frac{1}{2}L_{NN}-\dot{\mathcal{F}}\right) \delta N^{2}
+\frac{1}{2}\A \delta K^{2}+\B \delta K\delta N
+\C \delta K\delta _{1}\mathcal{R}
+{\cal D} \delta N\delta _{1}\mathcal{R} \nonumber \\
&&
+\mathcal{E}\delta _{2}\mathcal{R}+\frac{1}{2} 
\G \delta _{1}\mathcal{R}^{2} 
+L_{\s}\delta K_{\nu }^{\mu}\delta K_{\mu}^{\nu} 
+L_{\Z}\delta \mathcal{R}_{\nu}^{\mu} \delta 
\mathcal{R}_{\mu}^{\nu}\,,  \label{lagex}
\end{eqnarray}
where 
\begin{eqnarray}
\A &=&L_{KK}+4HL_{\s K} +4H^{2}L_{\s \s}\,,\\
\B &=&L_{KN}+2HL_{\s N}\,, \\
\C &=&L_{K\mathcal{R}}+2HL_{\s \mathcal{R}}
+\frac{1}{2} L_{\U}+HL_{K \U}+2H^{2}L_{\s \U}\,, \\
{\cal D} &=&L_{N\mathcal{R}}-\frac{1}{2}\dot{L_{\U}}
+HL_{N \U}\,, \\
\G &=&L_{\mathcal{R}\mathcal{R}}
+2HL_{\mathcal{R}\U}+H^{2}L_{\U \U}\,.
\end{eqnarray}
Then, we obtain the second-order Lagrangian density, as
\begin{eqnarray}
\hspace{-0.5cm}
\mathcal{L}_{2} &=&\delta \sqrt{h}
\left[ (\dot{\mathcal{F}}+L_{N})\delta N+ 
\mathcal{E}\delta _{1}\mathcal{R}\right]  \nonumber \\
\hspace{-0.5cm}
&&+a^{3}\biggl[\left( L_{N}+\frac{1}{2}L_{NN}\right) \delta N^{2}+\mathcal{E}
\delta _{2}\mathcal{R}+\frac{1}{2} \A \delta K^{2}+\B \delta
K\delta N+\C \delta K\delta _{1}\mathcal{R} \nonumber \\
& &~~~~~~~+({\cal D}+\mathcal{E})\delta N\delta _{1}\mathcal{R}
+\frac{1}{2} \G \delta _{1}\mathcal{R}^{2}
+L_{\s}\delta K_{\nu}^{\mu }\delta K_{\mu }^{\nu} 
+L_{\Z}\delta \mathcal{R}_{\nu}^{\mu} \delta 
\mathcal{R}_{\mu}^{\nu} \biggr]\,.  
\label{L2den}
\end{eqnarray}

For the gauge choice (\ref{gauge2}), the three-dimensional metric 
following from the metric (\ref{permet}) is 
$h_{ij}=a^2(t) e^{2\zeta} \delta_{ij}$. 
Then, several perturbed quantities appearing in Eq.~(\ref{L2den}) 
can be expressed as
\begin{eqnarray}
& & \delta \sqrt{h}=3a^{3}\zeta \,,\qquad \delta \mathcal{R}_{ij}=-\left( \delta
_{ij}\partial ^{2}\zeta +\partial _{i}\partial _{j}\zeta \right) \,,\nonumber \\
& & \delta _{1}\mathcal{R}=-4a^{-2}\partial ^{2}\zeta \,,\qquad \delta _{2} 
\mathcal{R}=-2a^{-2}\left[ (\partial \zeta )^{2}-4\zeta \partial ^{2}\zeta 
\right]\,,
\label{delh}
\end{eqnarray}
where $\partial ^{2}\zeta \equiv \partial _{i}\partial _{i}\zeta=
\sum_{i=1}^{3} \partial^2/\partial (x^i)^2$ and 
$(\partial \zeta )^{2}=(\partial_{i}\zeta )(\partial _{i}\zeta)=
\sum_{i=1}^{3} (\partial_i \zeta)^2$.

{}From Eq.~(\ref{Kdef}) the extrinsic curvature can be 
expressed in the form
\begin{equation}
K_{ij}=\frac{1}{2N} \left( \dot{h}_{ij}-N_{i|j}
-N_{j|i} \right)\,.
\end{equation}
For the perturbed metric (\ref{permet}), the first-order extrinsic 
curvature reads
\begin{equation}
\delta K_{j}^{i}=\left( \dot{\zeta}-H\delta N\right) \delta _{j}^{i} -\frac{1
}{2a^{2}}\delta ^{ik}(\partial _{k}N_{j}+\partial _{j}N_{k})\,,
\label{Kij}
\end{equation}
where we have used the fact that the Christoffel symbols 
$\Gamma _{ij}^{k}$ are the first-order perturbations 
for non-zero values of $k,i,j$. 
Since the shift $N_i$ is related to the metric perturbation $\psi$ via 
$N_i=\partial _{i}\psi$, the trace of $\delta K_{ij}$ can be expressed as
\begin{equation}
\delta K=3\left( \dot{\zeta}-H\delta N\right) 
-\frac{1}{a^{2}}\partial^{2}\psi \,. 
\label{delK2}
\end{equation}

On using the relations (\ref{delh}), (\ref{Kij}), and (\ref{delK2}), 
the second-order Lagrangian density (\ref{L2den}), up to 
boundary terms, reduces to
\begin{eqnarray}
\hspace{-0.3cm}
\mathcal{L}_2 &=& a^3 \biggl\{ \frac12 (2L_N+L_{NN}
+9\A H^2 -6\B H+6L_\s H^2) \delta N^2  \nonumber \\
& &~~~+\left[ (\B-3\A H-2L_\s H) 
\left( 3\dot{\zeta}-\frac{\partial^2 \psi}{a^2} \right) 
+4(3H \C-{\cal D}-\mathcal{E}) \frac{\partial^2 \zeta}{a^2} \right] \delta N 
\nonumber \\
\hspace{-0.3cm}
& &~~~-(3 \A+2L_\s) \dot{\zeta} \frac{\partial^2 \psi} 
{a^2} -12\C \dot{\zeta} \frac{\partial^2 \zeta}{a^2} 
+\left( \frac92 \A+3L_\s \right) \dot{\zeta}^2 +2\mathcal{E} 
\frac{(\partial \zeta)^2}{a^2}  \nonumber \\
\hspace{-0.3cm}
& &~~~+\frac12 ( \A+2L_\s ) \frac{(\partial^2 \psi)^2} 
{a^4} +4\C \frac{(\partial^2 \psi)(\partial^2 \zeta)}{a^4} 
+2(4\G+3L_{\Z}) \frac{(\partial^2 \zeta)^2}{a^4} \biggr\},
\label{L2exp}
\end{eqnarray}
where we have used the background equation (\ref{back3}) 
to eliminate the term $3a^3 (L_N+\dot{\mathcal{F}}) \zeta \delta N$.
Variations of the second-order action $S_{2}=\int d^{4}x\,\mathcal{L}_{2}$ with
respect to $\delta N$ and $\partial ^{2}\psi$ lead to the following 
Hamiltonian and momentum constraints, respectively:
\begin{eqnarray}
& &
\left[ 2L_{N}+L_{NN}-6H{\cal W}
-3H^{2}(3 \A+2L_{\s}) \right] \delta N \nonumber \\
& &
-{\cal W} \frac{\partial ^{2}\psi }{a^{2}}
+3{\cal W} \dot{\zeta} +4\left( 3H \C-{\cal D}-\mathcal{E} \right) 
\frac{\partial ^{2}\zeta}{a^{2}}=0\,,
\label{Hami} \\
& &
{\cal W} \delta N-(\A+2L_{\s}) 
\frac{\partial^{2}\psi}{a^{2}}+(3\A+2L_{\s})\dot{\zeta} 
-4\C \frac{\partial^{2}\zeta}{a^{2}}=0\,,
\label{momen}
\end{eqnarray}
where 
\begin{equation}
{\cal W} \equiv \B-3\A H-2L_{\s}H\,.
\end{equation}

{}From Eqs.~(\ref{Hami}) and (\ref{momen}) one can express $\delta N$ and 
$\partial ^{2}\psi /a^{2}$ in terms of $\dot{\zeta}$ and 
$\partial ^{2}\zeta /a^{2}$.
The last three terms in Eq.~(\ref{L2exp}) give rise to the equations 
of motion containing spatial derivatives higher than second order.
If we impose the three conditions 
\begin{eqnarray}
& & \A+2L_{\s} = 0\,,\label{elicon1}\\
& & \C = 0\,,\label{elicon2}\\
& & 4\G+3L_{\Z} = 0\,,\label{elicon3}
\end{eqnarray}
then such higher-order spatial derivatives are absent.
Under the conditions (\ref{elicon1})-(\ref{elicon3}), 
we obtain the following relations from 
Eqs.~(\ref{Hami}) and (\ref{momen}):
\begin{eqnarray}
\hspace{-0.5cm}
\frac{\partial^2 \psi}{a^2} &=&
\frac{3{\cal W}^2+4L_\s (2L_N+L_{NN}
-6H{\cal W}+12H^2L_\s)}{{\cal W}^2}
\dot{\zeta}-\frac{4({\cal D}+\mathcal{E})}{{\cal W}} 
\frac{\partial^2 \zeta}{a^2},\\
\hspace{-0.5cm}
\delta N &=& \frac{4L_\s}{{\cal W}} \dot{\zeta}\,,
\end{eqnarray}
where ${\cal W}=L_{KN}+2HL_{\s N}+4HL_{\s}$.
Substituting these relations into Eq.~(\ref{L2exp}), we find that the 
second-order Lagrangian density can be written in 
the form $\mathcal{L}_2=c_1(t) \dot{\zeta}^2
+c_2(t)\dot{\zeta} \partial^2 \zeta+c_3(t) (\partial \zeta)^2$, where 
$c_{1,2,3}(t)$ are time-dependent coefficients. 
After integration by parts, the term $c_2(t)\dot{\zeta} \partial^2 \zeta$ 
reduces to $\dot{c}_2(t) (\partial \zeta)^2/2$ up to a boundary term. 
Then, the second-order Lagrangian density reads \cite{Piazza,Gergely}
\begin{equation}
\mathcal{L}_2=a^3 Q_s \left[ \dot{\zeta}^2-\frac{c_s^2}{a^2}
(\partial \zeta)^2 \right]\,,
\label{perlag}
\end{equation}
where 
\begin{eqnarray}
Q_s &\equiv& \frac{2L_\s [3 \B^2+4L_\s (2L_N+L_{NN})]}
{{\cal W}^2}\,,
\label{Qsdef}\\
c_s^2 &\equiv& \frac{2}{Q_s} \left( \dot{\M}+H \M-\mathcal{E} \right)\,,
\label{csdef}
\end{eqnarray}
and
\begin{equation}
\M \equiv \frac{4L_\s ({\cal D}+{\cal E})}{{\cal W}}
=\frac{4L_\s}{{\cal W}} \left(
L_{\cal R}+L_{N{\cal R}}+HL_{N \U}
+\frac32 HL_\U \right)\,. 
\label{Mdef}
\end{equation}

Varying the action $S_2=\int d^4 x\,\mathcal{L}_2$ with respect to 
the curvature perturbation $\zeta$, 
we obtain the equation of motion for $\zeta$:
\begin{equation}
\frac{d}{dt} \left( a^3 Q_s \dot{\zeta} \right)-a Q_s c_s^2
\partial^2 \zeta=0\,. 
\label{pereq}
\end{equation}
This is the second-order equation of motion with a single scalar 
degree of freedom. Provided that the conditions (\ref{elicon1})-(\ref{elicon3}) 
are satisfied, the gravitational theory described by the action (\ref{action0}) 
does not involve derivatives higher than quadratic order at the level of {\it linear} 
cosmological perturbations. As we will see in Sec.~\ref{Hornsec}, Horndeski theory 
satisfies the conditions (\ref{elicon1})-(\ref{elicon3}).

While we have focused on scalar perturbations so far, we can also 
perform a similar expansion for tensor perturbations.
The three-dimensional metric including tensor modes $\gamma_{ij}$ 
can expressed as
\begin{equation}
h_{ij}=a^{2}(t)e^{2\zeta }\hat{h}_{ij}\,,\qquad \hat{h}_{ij}=\delta
_{ij}+\gamma _{ij}+\frac{1}{2}\gamma _{il}\gamma _{lj}\,,\qquad \mathrm{det}
\,\hat{h}=1\,,  \label{gra}
\end{equation}
where $\gamma_{ij}$ is traceless and divergence-free such that 
$\gamma _{ii}=\partial _{i}\gamma _{ij}=0$. 
We have introduced the second-order term 
$\gamma _{il}\gamma _{lj}/2$ for the simplification of 
calculations \cite{Maldacena}.
On using the property that tensor modes decouple from scalar modes, 
we substitute Eq.~(\ref{gra}) into the action (\ref{action0}) 
and then set scalar perturbations 0.
We note that tensor perturbations satisfy the relations 
$\delta K=0$, $\delta K_{ij}^{2}=\dot{\gamma}
_{ij}^{2}/4$, $\delta_{1}\mathcal{R}=0$, and 
$\delta _{2}\mathcal{R}=-(\partial _{k}\gamma _{ij})^{2}/(4a^2)$.
The second-order action for tensor perturbations reads
\begin{eqnarray}
S_2^{(h)} &=&
\int d^4 x\,a^{3}\left[ L_{\s} \left( \delta K_{\mu
}^{\nu }\delta K_{\nu }^{\mu }-\delta K^{2}\right) +
\mathcal{E}\delta _{2} 
\mathcal{R}\right] \nonumber \\
&=&\int d^4 x\,\frac{a^{3}}{4} \left[
L_{\s} \dot{\gamma}_{ij}^2-\frac{{\cal E}}{a^2}
(\partial_k \gamma_{ij})^2 \right]\,.
\label{tenlag}
\end{eqnarray}

One can express $\gamma_{ij}$ in terms of two polarization modes, as 
$\gamma_{ij}=h_+e_{ij}^{+}+h_{\times}e_{ij}^{\times}$.
In Fourier space, the transverse and traceless tensors 
$e_{ij}^{+}$ and $e_{ij}^{\times}$
satisfy the normalization condition
$e_{ij} (\bm{k})\,e_{ij} (-\bm{k})^{*}=2$ for each 
polarization ($\bm{k}$ is a comoving wavenumber), 
whereas $e^{+}_{ij} (\bm{k})\,e^{\times}_{ij}(-\bm{k})^{*}=0$.
The second-order Lagrangian (\ref{tenlag}) can be written 
as the sum of two polarizations, as 
\begin{equation}
S_2^{(h)} =
\sum_{\lambda=+,\times}
\int d^4 x~a^{3} Q_t \left[ \dot{h}_{\lambda}^2
-\frac{c_t^2}{a^2} (\partial h_{\lambda})^2 \right]\,,
\label{tenlag2}
\end{equation}
where
\begin{eqnarray}
Q_t &\equiv& \frac{L_\s}{2}\,,\\
c_t^2 &\equiv& \frac{\mathcal{E}}{L_\s}\,.
\end{eqnarray}
Each mode $h_{\lambda}$ ($\lambda=+,\times$) obeys 
the second-order equation of motion
\begin{equation}
\frac{d}{dt} \left( a^3 Q_t \dot{h}_{\lambda} \right)-a Q_t c_t^2
\partial^2 h_{\lambda}=0\,. 
\label{tenpereq}
\end{equation}

In order to avoid the appearance of ghosts, the coefficient in 
front of the term $\dot{h}_{\lambda}$ needs to be positive 
and hence $Q_t>0$. The small-scale instability associated 
with the Laplacian term $c_t^2 \partial^2 h_{\lambda}$ is absent 
for $c_t^2>0$. Then, the conditions for avoidance of the ghost 
and the Laplacian instability associated with tensor perturbations 
are given, respectively, by \cite{Piazza,Gergely}
\begin{eqnarray}
& &L_\s>0\,,\label{ghost1}\\
& &\mathcal{E}=L_{\cal R}+\frac12 \dot{L_\U}+
\frac32 H L_{\U}>0\,.\label{ins1}
\end{eqnarray}
Similarly, the ghost and the Laplacian instability of scalar perturbations
can be avoided for $Q_s>0$ and $c_s^2>0$, respectively, i.e., 
\begin{eqnarray}
& &3 \left(L_{KN}+2HL_{\s N} \right)^2+
4L_\s (2L_N+L_{NN})>0\,,\label{ghost2}\\
& &\dot{\M}+H \M-{\cal E}
>0\,,\label{ins2}
\end{eqnarray}
where we have used the condition (\ref{ghost1}).
The four conditions (\ref{ghost1})-(\ref{ins2}) need to be satisfied 
for theoretical consistency.

\section{Inflationary power spectra}	
\label{infsec}

The scalar degree of freedom discussed in the previous section 
can give rise to inflation in the early Universe. 
Moreover, the curvature perturbation $\zeta$ generated during 
inflation can be responsible for the origin of 
observed CMB temperature anisotropies \cite{oldper}.
The tensor perturbation not only contributes to the CMB power 
spectrum but also leaves an imprint for the B-mode 
polarization of photons.

In this section we derive the inflationary power spectra of 
scalar and tensor perturbations for the general action (\ref{action0}). 
We focus on the theory satisfying the conditions (\ref{elicon1})-(\ref{elicon3}).
In this case, the equations of linear cosmological perturbations 
do not involve time and spatial derivatives higher than second order.
Since the Hubble parameter $H$ is nearly constant during inflation, 
the terms that do not contain the scale factor $a$ slowly vary in time. 
Let us then assume that variations of the terms
$Q_s$, $Q_t$, $c_s$, and $c_t$ 
are small, such that the quantities
\begin{equation}
\delta_{Q_s} \equiv \frac{\dot{Q}_s}{HQ_s}\,,\quad
\delta_{Q_t} \equiv \frac{\dot{Q}_t}{HQ_t}\,,\quad
\delta_{c_s}  \equiv \frac{\dot{c}_s}{Hc_s}\,,\quad
\delta_{c_t}  \equiv \frac{\dot{c}_t}{Hc_t}
\label{Qcdef}
\end{equation}
are much smaller than unity.

We first study the evolution of the curvature perturbation 
$\zeta$ during inflation. In doing so, we express $\zeta$ 
in Fourier space, as 
\begin{equation}
\zeta (\tau,{\bm{x}})
= \frac{1}{(2\pi)^{3}}\int d^{3}{\bm{k}}\, 
\hat{\zeta} (\tau,{\bm{k}})e^{i{\bm{k}}\cdot{\bm{x}}}\,,
\label{RFourier}
\end{equation}
where 
\begin{equation}
\hat{\zeta}(\tau,{\bm{k}})=u(\tau,{\bm{k}})a({\bm{k}})
+u^{*}(\tau,{-\bm{k}})a^{\dagger}(-{\bm{k}})\,.
\end{equation}
Here, $\tau \equiv \int a^{-1}\,dt$ is the conformal time, 
${\bm k}$ is the comoving wavenumber, 
$a({\bm{k}})$ and $a^{\dagger}({\bm{k}})$ are the annihilation
and creation operators, respectively, satisfying the commutation relations
\begin{eqnarray}
& &
\left[a({\bm{k}}_{1}),a^{\dagger}({\bm{k}}_{2})\right]
=(2\pi)^{3}\delta^{(3)}({\bm{k}}_{1}-{\bm{k}}_{2})\,,\nonumber \\
& &
\left[a({\bm{k}}_{1}),a({\bm{k}}_{2})\right]
=\left[a^{\dagger}({\bm{k}}_{1}),a^{\dagger}({\bm{k}}_{2})\right]=0\,.
\end{eqnarray}
On the de Sitter background where $H$ is constant, we have 
$a \propto e^{Ht}$ and hence $\tau=-1/(aH)$. 
Here, we have set the integration constant 0, such that the asymptotic
past corresponds to $\tau \to -\infty$.

Using the equation of motion (\ref{pereq}) for $\zeta$, 
we find that each Fourier mode $u$ obeys 
\begin{equation}
\ddot{u}+\frac{(a^{3}Q_s)^{\cdot}}{a^{3}Q_s}\dot{u}
+c_{s}^{2}\frac{k^{2}}{a^{2}}u=0\,.
\label{ueq}
\end{equation}
For large $k$, the second term on the l.h.s. of  
Eq.~(\ref{ueq}) is negligible relative to the third one, 
so that the field $u$ oscillates according to the approximate 
equation $\ddot{u}+c_s^2(k^2/a^2)u \simeq 0$. 
After the onset of inflation, the $c_s^2(k^2/a^2)u$ term starts to 
decrease quickly. Since the second term on the l.h.s. of Eq.~(\ref{ueq})
is of the order of $H^2 u$, the third term becomes negligible 
relative to the other terms for $c_sk<aH$.
In the large-scale limit ($k \to 0$), the solution 
to Eq.~(\ref{ueq}) is given by 
\begin{equation}
u=c_1+c_2 \int \frac{1}{a^3 Q_s}\,dt\,,
\end{equation}
where $c_1$ and $c_2$ are integration constants. 
As long as the variable $Q_s$ changes slowly
in time, $u$ approaches a constant value $c_1$.
The field $u$ starts to be frozen once the perturbations 
with the wavenumber $k$ cross 
$c_sk=aH$ \cite{oldper,kinfper,BTW}.

We recall that the second-order Lagrangian for the curvature 
perturbation $\zeta$ is given by Eq.~(\ref{perlag}).
Introducing a rescaled field $v=zu$ with $z=a\sqrt{2Q_s}$,
the kinetic term in the second-order action $S_2=\int d^4x \,{\cal L}_2$
can be rewritten as $\int d\tau d^3 x\,v'^2/2$, 
where a  prime represents a derivative with respect to $\tau$. 
This means that $v$ is a canonical field that should 
be quantized \cite{Muka,Tsujinon}.
Equation (\ref{ueq}) can be written as 
\begin{equation}
v''+\left(c_{s}^{2}k^{2}-\frac{z''}{z}\right)v=0\,.
\label{veq}
\end{equation}
On the de Sitter background with a slow variation of 
the quantity $Q_s$, we can approximate $z''/z \simeq 2/\tau^{2}$. 
In the asymptotic past ($k\tau\to-\infty$), we choose 
the Bunch-Davies vacuum characterized by 
the mode function $v=e^{-ic_{s}k\tau}/\sqrt{2c_{s}k}$.
Then the solution to Eq.~(\ref{veq}) 
is given by \cite{kinfper,XGao,Tsujinon,Tsujinon2}
\begin{equation}
u (\tau, k)=\frac{i\,H\, e^{-ic_{s}k\tau}}
{2(c_{s}k)^{3/2}\sqrt{Q_s}}\,(1+ic_{s}k\tau)\,.
\label{usol}
\end{equation}
The deviation from the exact de Sitter background gives 
rise to a small modification to the solution (\ref{usol}), 
but this difference appears as a next-order 
slow-roll correction to the power spectrum \cite{Chen,Hornshape}.
 
In the regime $c_sk \ll aH$, the two-point correlation function 
of $\zeta$ is given by the vacuum expectation value
$\langle 0| \hat{\zeta} (\tau, {\bm k}_1) \hat{\zeta} 
(\tau,{\bm k}_2) | 0 \rangle$ at $\tau \approx 0$. 
We define the scalar power spectrum 
${\cal P}_{\zeta} (k_1)$, as 
\begin{equation}
\langle 0| \hat{\zeta} (0,{\bm k}_1) \hat{\zeta} (0,{\bm k}_2) | 0 \rangle
=\frac{2\pi^2}{k_1^3} {\cal P}_{\zeta} (k_1)\,
(2\pi)^3 \delta^{(3)} ({\bm k}_1+{\bm k}_2)\,.
\label{powerdef}
\end{equation}
Using the solution (\ref{usol}) in the $\tau \to 0$ limit,
it follows that 
\begin{equation}
{\cal P}_{\zeta}=\frac{H^2}{8\pi^2 Q_s c_s^3}\,.
\label{scalarpower}
\end{equation}
Since the curvature perturbation soon approaches a constant 
for $c_s k <aH$, it is a good approximation to evaluate
the power spectrum (\ref{scalarpower}) at
$c_s k=aH$ during inflation.
From the Planck data, the scalar amplitude is constrained as
${\cal P}_{\zeta} \simeq  2.2 \times 10^{-9}$ at the pivot 
wavenumber $k_0=0.002$ Mpc$^{-1}$ \cite{Planck}.

The spectral index of ${\cal P}_\zeta$ is defined by 
\begin{eqnarray}
n_{s}-1 \equiv \frac{d \ln {\cal P}_{\zeta}}
{d \ln k}\bigg|_{c_sk=aH}
=-2\epsilon-\delta_{Q_s}-3\delta_{c_s}\,,
\label{ns}
\end{eqnarray}
where $\delta_{Q_s}$ and $\delta_{c_s}$ are given
by Eq.~(\ref{Qcdef}), and 
\begin{equation}
\epsilon \equiv -\frac{\dot{H}}{H^2}\,.
\end{equation}
The slow-roll parameter $\epsilon$ is much smaller than 1 
on the quasi de Sitter background. 
Given that the variations of $H$ and $c_s$ are small during inflation, 
we can approximate the variation of $\ln k$ at $c_s k=aH$, as $d \ln k=d \ln a=H dt$.
Since we are considering the situation with $|\delta_{Q_s}| \ll 1$ 
and $|\delta_{c_s}| \ll 1$, the power spectrum is close to 
scale-invariant ($n_s \simeq 1$).

We also define the running of the spectral index, as
\begin{equation}
\alpha_s \equiv \frac{dn_s}{d \ln k}\biggr|_{c_{s}k=aH}\,,
\end{equation}
which is of the oder of $\epsilon^2$ from Eq.~(\ref{ns}).
With the prior $\alpha_s=0$, the scalar spectral index 
is constrained as $n_s=0.9603 \pm 0.0073$ at 68\,\%\,confidence level (CL) from 
the Planck data \cite{Planck}. Since $\epsilon$ is at most of the order $10^{-2}$, 
it is a good approximation to neglect the running $\alpha_s$ 
in standard slow-roll inflation.

Let us also derive the spectrum of gravitational waves generated 
during inflation. 
The second-order action for tensor perturbations is given by 
Eq.~(\ref{tenlag2}), where $h_{\lambda}$ obeys Eq.~(\ref{tenpereq}).
A canonical field associated with $h_{\lambda}$ ($\lambda=+, \times$) 
corresponds to $v_t=z_t h_{\lambda}$ and $z_t=a\sqrt{2Q_t}$. 
Following a same procedure as that for scalar perturbations, 
the solution to the Fourier-transformed mode 
$h_{\lambda}$, which recovers the Bunch-Davies 
vacuum in the asymptotic past, reads
\begin{equation}
h_{\lambda} (\tau,k)=\frac{i\,H\, e^{-ic_{t}k\tau}}
{2(c_{t}k)^{3/2}\sqrt{Q_t}}\,(1+ic_{t}k\tau)\,.
\label{hsol}
\end{equation}
This solution approaches $h_{\lambda} \to i H/[2(c_tk)^{3/2} \sqrt{Q_t}]$
in the $\tau \to 0$ limit.

We also define the tensor power spectrum ${\cal P}_{h}$ in 
a similar way to (\ref{powerdef}).
According to the chosen normalization for the tensors 
$e^{\lambda}_{ij}$ explained in Sec.~\ref{secondsec}, 
we obtain ${\cal P}_h=4 \cdot k^3 |h_{\lambda} (0,k)|^2/(2\pi^2)$, 
where $h_{\lambda}(0,k)=i H/[2(c_tk)^{3/2} \sqrt{Q_t}]$.
It then follows that 
\begin{equation}
{\cal P}_h=\frac{H^2}{2\pi^2 Q_t c_t^3}\,.
\label{tensorpower}
\end{equation}
The tensor spectral index, which is evaluated 
at $c_tk=aH$, reads
\begin{equation}
n_t \equiv \frac{d \ln {\cal P}_h}
{d \ln k}\bigg|_{c_t k=aH}
=-2\epsilon-\delta_{Q_t}-3\delta_{c_t}\,,
\label{nT}
\end{equation}
where $\delta_{Q_t}$ and $\delta_{c_t}$ are given by Eq.~(\ref{Qcdef}).
The tensor power spectrum is close to scale-invariant ($n_t \simeq 0$)
provided that $\epsilon \ll 1$, $|\delta_{Q_t}| \ll 1$,
and $|\delta_{c_t}| \ll 1$. 
The difference between the scalar and tensor spectral 
indices comes from the difference between $(Q_s,c_s)$ 
and $(Q_t,c_t)$.

For those times before the end of inflation ($\epsilon \ll 1$)
when both ${\cal P}_{\zeta}$ and ${\cal P}_{h}$ are 
approximately constant, the tensor-to-scalar ratio 
can be estimated as
\begin{equation}
r \equiv \frac{{\cal P}_h}{{\cal P}_{\zeta}}
=4\frac{Q_s c_s^3}{Q_t c_t^3}\,.
\label{rfi}
\end{equation}
The Planck data \cite{Planck}, combined with the WMAP large-angle 
polarization measurement \cite{WMAP9} and ACT/SPT 
temperature data \cite{ACT}, 
showed that $r$ is constrained as $r<0.11$ (95\,\%\,CL). 
Recently, the Background Imaging of Cosmic Extragalactic 
Polarization (BICEP2) group \cite{biceps2} reported the 
first evidence for the primordial B-mode polarization 
of CMB photons and they derived the bound 
$r=0.20^{+0.07}_{-0.05}$ (68\,\%\,CL) 
with $r=0$ disfavored at 7$\sigma$.
There is a tension between the data of Planck and BICEP2, but 
future measurements of the B-mode 
polarization will place more precise bounds on $r$.

The inflationary scalar and tensor power spectra (\ref{scalarpower}) and (\ref{tensorpower}) 
are valid for general theories given by the action (\ref{action0}), 
provided that the conditions (\ref{elicon1})-(\ref{elicon3}) are satisfied.
The quantities like $Q_s$ and $c_s^2$ are written in terms of 
the partial derivatives of $L$ with respect to the ADM variables 
such as $K$ and $N$.
For a given theory, we need to express the Lagrangian $L$ in 
terms of the three-dimensional quantities and the lapse $N$ 
to derive concrete forms of the inflationary power spectra. 
In the next section, we will perform this procedure for the 
most general scalar-tensor theories with second-order 
equations of motion.

\section{Horndeski theory}	
\label{Hornsec}

%
\subsection{The Lagrangian of Horndeski theory}
\label{conmodel}

In this section we apply the EFT formalism advocated in Secs.~\ref{actionsec} 
and \ref{secondsec} to the most general scalar-tensor theories with second-order 
equations of motion--Horndeski theory \cite{Horndeski}.
This theory is described by the action 
$S=\int d^4 x\,\sqrt{-g}\,L$, with the Lagrangian \cite{Deffayet}
\begin{equation}
L=\sum_{i=2}^{5} L_{i}\,,\label{Lagsum}
\end{equation}
where 
\begin{eqnarray}
L_{2} & = & G_2(\phi,X),\label{eachlag2}\\
L_{3} & = &  G_{3}(\phi,X)\square\phi,\\
L_{4} & = & G_{4}(\phi,X)\, R-2G_{4X}(\phi,X)\left[ (\square \phi)^{2}
-\phi^{;\mu \nu }\phi _{;\mu \nu} \right] \,, \label{L4lag}\\
L_{5} &=&G_{5}(\phi,X)G_{\mu \nu }\phi ^{;\mu \nu} \nonumber \\
& &+\frac{1}{3}G_{5X}(\phi,X)
\left[ (\square \phi )^{3}-3(\square \phi )\,\phi _{;\mu \nu }\phi ^{;\mu
\nu }+2\phi _{;\mu \nu }\phi ^{;\mu \sigma }{\phi ^{;\nu}}_{;\sigma }
\right]\,. 
\label{L5lag}
\end{eqnarray}
Here $G_{i}$ ($i=2,3,4,5$) are functions in terms of a scalar
field $\phi$ and its kinetic energy 
$X=g^{\mu \nu}\partial_{\mu} \phi \partial_{\nu} \phi$
with the partial derivatives $G_{iX} \equiv\partial G_{i}/\partial X$ 
and $G_{i\phi} \equiv \partial G_{i}/\partial \phi$,  
$R$ is the Ricci scalar, and $G_{\mu\nu}$ is the Einstein tensor.
In 1973, Horndeski derived the Lagrangian of the most general 
scalar-tensor theories in a different form \cite{Horndeski}, but 
as shown in Ref.~\cite{KYY}, it is equivalent to the above form. 
The Horndeski's paper\footnote{When Horndeski wrote this paper, 
he was the PhD student of David Lovelock. In 1981, 
he was taking a sabbatical year in Netherlands as  
a tenured professor of applied mathematics at the University of Waterloo. 
When he saw a van Gogh exhibition, he was deeply moved. 
He stated ``I was never that interested in art. Then I stumbled onto van Gogh. 
I never knew art could be like that. I had always thought of it as very representational
and not very interesting. But then I thought, 
`This is something I eventually want to do.' 
When I saw van Gogh I was sure I could paint.'' 
After this, Horndeski left physics and became an artist.} has not been recognized 
much for a long time, but it was 
revived recently in connection to covariant Galileons \cite{Nicolis,cova} 
and generalized Galileon 
theories \cite{Char,Deffayet}.

The Lagrangian (\ref{Lagsum}) covers a wide variety of 
gravitational theories listed below.
\begin{itemize}
\item (1) General Relativity with a minimally coupled scalar field
\vspace{0.2cm}

The minimally coupled scalar-field theory (\ref{klag}) 
is characterized by the functions \cite{kinf}
\begin{equation}
G_2=P(\phi,X)\,,\qquad G_3=0\,,\qquad
G_4=M_{\rm pl}^2/2\,,\qquad G_5=0\,.
\end{equation}
The canonical scalar field with a potential $V(\phi)$ corresponds to 
the particular choice 
\begin{equation}
G_2=-X/2-V(\phi)\,.
\end{equation}
\item (2) Brans-Dicke theory
\vspace{0.2cm}

The Lagrangian of Brans-Dicke (BD) theory is given by  
\begin{equation}
G_2=-\frac{M_{\rm pl}\omega_{\rm BD} X}{2\phi}-V(\phi)\,,
\qquad G_3=0\,,\qquad
G_4=\frac{1}{2}M_{\rm pl}\phi\,,\qquad G_5=0\,,
\label{BDaction}
\end{equation}
where $\omega_{\rm BD}$ is the so-called BD parameter.
In the original BD theory \cite{Brans},  the field potential $V(\phi)$ is absent.
Dilaton gravity \cite{dilaton} corresponds to $\omega_{\rm BD}=-1$.

\vspace{0.2cm}
\item (3) $f(R)$ gravity
\vspace{0.2cm}

This theory is characterized by the action
\begin{equation}
S=\int d^4x \sqrt{-g}\,\frac{M_{\rm pl}^2}{2}f(R)\,,
\label{fRaction}
\end{equation}
where $f(R)$ is an arbitrary function of the Ricci scalar $R$.
The metric $f(R)$ gravity corresponds to the case in which 
the action (\ref{fRaction}) is varied with respect to $g_{\mu \nu}$.
This can be accommodated by the Lagrangian (\ref{Lagsum}) 
for the choice
\begin{equation}
G_2=-\frac{M_{\rm pl}^2}{2} (RF-f),\qquad
G_3=0\,,\qquad G_4=\frac{1}{2}M_{\rm pl}^2F\,,\qquad G_5=0\,,
\label{fRcase}
\end{equation}
where $F \equiv \partial f/\partial R$. 
There is a scalar degree of freedom $\phi=M_{\rm pl}F(R)$ 
with a gravitational origin. 
Comparing Eq.~(\ref{BDaction}) with Eq.~(\ref{fRcase}), we find that 
metric $f(R)$ gravity is equivalent to BD theory with 
$\omega_{\rm BD}=0$ and the potential 
$V=(M_{\rm pl}^2/2) (RF-f)$.

In the Palatini formalism where the metric $g_{\mu \nu}$ and the connection 
$\Gamma^{\alpha}_{\beta \gamma}$ are treated as independent variables, 
the Ricci scalar is different from that in metric $f(R)$ gravity. 
The Palatini $f(R)$ gravity is equivalent to BD theory with the 
parameter $\omega_{\rm BD}=-3/2$ \cite{moreview}.

\vspace{0.2cm}
\item (4) Non-minimally coupled theory
\vspace{0.2cm}

This theory is described by the functions 
\begin{equation}
G_2=\omega (\phi)X-V(\phi)\,,\qquad G_3=0\,,
\qquad G_4=\frac{M_{\rm pl}^2}{2}-\frac12 \xi \phi^2\,,
\qquad G_5=0\,.
\end{equation}
where $\omega(\phi)$ and $V(\phi)$ are functions 
of $\phi$. Higgs inflation \cite{Higgs} corresponds to 
a canonical field ($\omega(\phi)=-1/2$) 
with the potential $V(\phi)=(\lambda/4)(\phi^2-v^2)^2$ 
(see also Refs.~\cite{Higgsearly}).

\vspace{0.2cm}
\item (5) Covariant Galileons
\vspace{0.2cm}

The covariant Galileons \cite{cova}, in the absence of 
the field potential, are described by the functions
\begin{equation}
G_2=c_2X\,,\qquad G_3=c_3X\,,\qquad
G_4=\frac{M_{\rm pl}^2}{2}+c_4 X^2\,,\qquad
G_5=c_5 X^2\,,
\end{equation}
where $c_{i}$ ($i=2,3,4,5$) are constants. The field equations of motion 
are invariant under the Galilean transformation 
$\partial_{\mu} \phi \to \partial_{\mu} \phi+b_{\mu}$
in the limit of Minkowski space-time \cite{Nicolis}.

\vspace{0.2cm}
\item (6) Derivative couplings
\vspace{0.2cm}

A scalar field whose derivative couples to 
the Einstein tensor in the form 
$G_{\mu \nu} \partial^{\mu} \phi \partial^{\nu} \phi$ \cite{Amendola93,Germani}
corresponds to the choice
\begin{equation}
G_2=-X/2-V(\phi)\,,\qquad G_3=0\,,\qquad
G_4=0\,,\qquad
G_5=c \phi\,,
\end{equation}
where $c$ is a constant and $V(\phi)$ is the field potential.
In fact, integration of the term $c\phi G_{\mu \nu} \phi^{;\mu \nu}$ by parts 
gives rise to the coupling $-cG_{\mu \nu}\partial^{\mu} \phi \partial^{\nu} \phi$.

\vspace{0.2cm}
\item (7) Gauss-Bonnet couplings 
\vspace{0.2cm}

The Gauss-Bonnet couplings of the from $-\xi(\phi)R_{\rm GB}^2$, 
where $R_{\rm GB}^2=R^2-4R_{\alpha \beta}R^{\alpha \beta}
+R_{\alpha \beta \gamma \delta}R^{\alpha \beta \gamma \delta}$, 
can be accommodated by the choice \cite{KYY}
\begin{eqnarray}
&&
G_2=-2\xi^{(4)} (\phi) X^2 [3-\ln (-X/2)]\,,\quad
G_3=2\xi^{(3)} (\phi) X [7-3\ln (-X/2)]\,,\nonumber \\
& &
G_4=2\xi^{(2)}(\phi)X [2-\ln (-X/2)]\,,\qquad~
G_5=4\xi^{(1)} (\phi) \ln (-X/2)\,,
\end{eqnarray}
where $\xi^{(n)}(\phi)=\partial^n \xi (\phi)/\partial \phi^n$.
\end{itemize}

\subsection{Horndeski Lagrangian in terms of ADM variables}

Let us express the Horndeski Lagrangians (\ref{eachlag2})-(\ref{L5lag}) 
in terms of the lapse $N$ and the three-dimensional quantities 
introduced in Sec.~\ref{actionsec}.
In unitary gauge, the unit vector $n_{\mu}$ orthogonal to 
the constant $\phi$-hypersurface is given by \cite{Piazza}
\begin{equation}
n_{\mu}=-\gamma \phi_{;\mu}\,,\qquad
\gamma=\frac{1}{\sqrt{-X}}\,.
\label{nmu}
\end{equation}
Taking the covariant derivative of Eq.~(\ref{nmu}) and using the relation 
(\ref{Rre}), we obtain 
\begin{equation}
\phi_{;\mu \nu} =-\frac{1}{\gamma}\left( K_{\mu \nu }-n_{\mu }a_{\nu
}-n_{\nu }a_{\mu}\right) +\frac{\gamma ^{2}}{2} \phi^{;\sigma}X_{;\sigma}
n_{\mu }n_{\nu}\,.
\label{phimunu}
\end{equation}
The trace of Eq.~(\ref{phimunu}) gives
\begin{equation}
\square \phi=-\frac{1}{\gamma}K+
\frac{\phi^{;\sigma}X_{;\sigma}}{2X}\,.
\label{sqphi}
\end{equation}

First of all, the Lagrangian $L_2$ depends on 
$N$ through the field kinetic energy, i.e., 
\begin{equation}
L_2=G_2(\phi, X(N))\,.
\label{L2three}
\end{equation}
On using the property $X(N)=-\dot{\phi}^2/N^2$ on the flat
FLRW background, the quantity like $L_{2N}$
can be evaluated as $L_{2N}=2\dot{\phi}^2G_{2X}$.

For the computation of $L_3=G_3 \square \phi$, 
it is convenient to introduce an 
auxiliary function $F_3(\phi,X)$, as 
\begin{equation}
G_3=F_3+2XF_{3X}\,.
\end{equation}
After integration by parts, the term $F_3 \square \phi$ 
reduces to $-(F_{3\phi} \phi_{;\mu}+F_{3X}X_{;\mu})\phi^{;\mu}$ 
up to a boundary term. On using the relation (\ref{sqphi}) 
for the term $2XF_{3X} \square \phi$, it follows that 
\begin{equation}
L_3=2(-X)^{3/2}F_{3X}K-XF_{3\phi}\,.
\label{L3three}
\end{equation}
Although the auxiliary function $F_3$ is present
in the expression of $L_3$, the combination of quantities 
appearing in the background and linear perturbation equations of 
motion can be expressed in terms of $G_3$.

Substituting Eqs.~(\ref{phimunu}) and (\ref{sqphi}) 
into Eq.~(\ref{L4lag}), the term $L_4$ reads
\begin{equation}
L_4=G_4R+2XG_{4X} (K^2-{\cal S})
+2G_{4X}X_{;\mu} (Kn^{\mu}-a^{\mu})\,,
\label{L4ex}
\end{equation}
where we have used the property $a_{\mu}=-h_{\mu}^{\nu}X_{;\nu}/(2X)$.
Substituting Eq.~(\ref{Rre}) into Eq.~(\ref{L4ex}) and 
employing the relations $G_{4X}X_{;\mu}=G_{4;\mu}
+\gamma^{-1}G_{4\phi}n_{\mu}$ and $n_{\mu}a^{\mu}=0$, 
we obtain 
\begin{equation}
L_4=G_4{\cal R}+(2XG_{4X}-G_4)(K^2-{\cal S})
-2\sqrt{-X}G_{4\phi}K\,.
\label{L4three}
\end{equation}

The Lagrangian $L_5$ is most complicated to be dealt with.
We refer readers to Ref.~\cite{Piazza} for detailed calculations.
Introducing an auxiliary function $F_5 (\phi,X)$ such that 
\begin{equation}
G_{5X} \equiv \frac{F_5}{2X}+F_{5X}\,,
\end{equation}
the final expression of $L_5$ is given by 
\begin{eqnarray}
L_{5} &=&
\sqrt{-X}F_{5}\left( \frac{1}{2}K\mathcal{R}- 
{\cal U} \right) -H(-X)^{3/2}G_{5X}(2H^{2}-2KH+K^{2}-{\cal S}) \nonumber \\
& &+\frac{1}{2}X(G_{5\phi}-F_{5\phi})\mathcal{R}
+\frac{1}{2}XG_{5\phi} (K^{2}-{\cal S})\,,
\label{L5three}
\end{eqnarray}
which is valid up to quadratic order in the perturbations.

Summing up the contributions (\ref{L2three}), (\ref{L3three}), (\ref{L4three}), 
and (\ref{L5three}), the Lagrangian (\ref{Lagsum})
can be expressed as
\begin{eqnarray}
L &=& G_2+2(-X)^{3/2}F_{3X}K-XF_{3\phi} \nonumber \\
& &
+G_4{\cal R}+(2XG_{4X}-G_4)(K^2-{\cal S})
-2\sqrt{-X}G_{4\phi}K \nonumber \\
& &
+\sqrt{-X}F_{5}\left( \frac{1}{2}K\mathcal{R}- 
{\cal U} \right) -H(-X)^{3/2}G_{5X}
(2H^{2}-2KH+K^{2}-{\cal S}) 
\nonumber \\
& &
+\frac{1}{2}X(G_{5\phi}-F_{5\phi})\mathcal{R}
+\frac{1}{2}XG_{5\phi} (K^{2}-{\cal S})\,,
\label{Ltotal}
\end{eqnarray}
where $G_{2,3,4,5}$ and $F_{3,5}$ are functions of 
$\phi$ and $X(N)$. 
The Lagrangian (\ref{Ltotal}) depends on $N$, $K$, 
${\cal S}$, ${\cal R}$, $\U$, but not on $\Z$.
We evaluate the partial derivatives of the 
Lagrangian (\ref{Ltotal}) with respect to $N$, $K$ e.t.c.
and finally set $N=1$, $K=3H$, 
${\cal S}=3H^2$, ${\cal R}=0$, $\U=0$.

Among the terms appearing in Eqs.~(\ref{elicon1})-(\ref{elicon3}), 
the non-vanishing ones are given by 
\begin{eqnarray}
L_{KK} &=& -2L_\s=2(2X G_{4X}-G_4)-2H(-X)^{3/2}G_{5X}+XG_{5\phi}\,,\\
L_{K{\cal R}}&=&-\frac12 L_\U=\frac12 \sqrt{-X}F_5\,,
\end{eqnarray}
so that all the three conditions (\ref{elicon1})-(\ref{elicon3}) 
are satisfied. In Horndeski theory, there are no spatial
derivatives higher than second order.

\subsection{Conditions for the avoidance of ghosts and 
Laplacian instabilities}

The conditions (\ref{ghost1}) and (\ref{ins1}) for avoiding 
the ghost and the Laplacian instability of tensor perturbations 
translate to 
\begin{eqnarray}
L_{\s} &=& G_{4}-2XG_{4X}-H\dot{\phi}XG_{5X} 
-\frac{1}{2}XG_{5\phi}>0\,, \label{Ls} \\
\mathcal{E} &=&G_{4}+\frac{1}{2}XG_{5\phi}-XG_{5X}\ddot{\phi}>0\,,
\label{calE}
\end{eqnarray}
respectively.
In the presence of the terms $G_4(X)$ and $G_5(\phi,X)$, 
the tensor propagation speed square $c_t^2=\mathcal{E}/L_\s$
is generally different from 1.

On using the properties ${\cal B}=L_{KN}+2HL_{\s N}$ and
${\cal W}=L_{KN}+2HL_{\s N}+4HL_{\s}$, the quantity $Q_s$ 
in Eq.~(\ref{Qsdef}) can be expressed as 
\begin{equation}
Q_s=\frac{2L_\s}{3{\cal W}^2} 
\left( 9{\cal W}^2+8L_\s w \right)\,,
\label{Qsho}
\end{equation}
where\footnote{The four quantities $w_{1,2,3,4}$ 
introduced in Ref.~\cite{Tsujinon2} are related to $L_{\s}$, 
${\cal W}$, $w$, and ${\cal E}$, as 
$w_1=2L_\s$, $w_2={\cal W}$, $w_3=w$, and $w_4=2{\cal E}$.} 
\begin{eqnarray}
\hspace{-0.1cm}
w &\equiv &3L_{N}+3L_{NN}/2-9H(L_{KN}+2HL_{\s N}) 
-18L_{\s}H^{2}  \nonumber \\
&=&-18H^{2}G_{4}+3(XG_{2X}+2X^{2}G_{2XX})-18H\dot{\phi}
(2XG_{3X}+X^{2}G_{3XX}) \nonumber \\
&& -3X(G_{3\phi}+XG_{3\phi X}) 
+18H^{2}(7XG_{4X}+16X^{2}G_{4XX}+4X^{3}G_{4XXX}) \nonumber \\
&& -18H\dot{\phi}(G_{4\phi}+5XG_{4\phi X}+2X^{2}G_{4\phi XX})  
+6H^{3}\dot{\phi}(15XG_{5X}+13X^{2}G_{5XX} \nonumber \\
&& +2X^{3}G_{5XXX})
+9H^{2}X(6G_{5\phi}+9XG_{5\phi X}+2X^{2}G_{5\phi XX})\,, \\
\hspace{-0.1cm}
{\cal W} &=&4HG_{4}+2\dot{\phi}XG_{3X}-16H(XG_{4X}+X^{2}G_{4XX}) 
+2\dot{\phi}(G_{4\phi}+2XG_{4\phi X})\nonumber \\
&&
-2H^{2}\dot{\phi}(5XG_{5X}+2X^{2}G_{5XX}) 
-2HX(3G_{5\phi}+2XG_{5\phi X})\,.  \label{Wre}
\end{eqnarray}
Taking into account the requirement (\ref{Ls}), 
the no-ghost condition for scalar perturbations reads
\begin{equation}
9{\cal W}^2+8L_\s w>0\,.
\end{equation}

In Horndeski theory (\ref{Ltotal}), we notice that there is 
the following relation 
\begin{equation}
L_{\s}={\cal D}+\mathcal{E}=L_{\mathcal{R}} 
+L_{N\mathcal{R}}+\frac32 HL_{\U}+HL_{N \U}\,,
\end{equation}
so that the quantity (\ref{Mdef}) reduces to 
\begin{equation}
{\cal M}=\frac{4L_\s^2}{{\cal W}}\,.
\end{equation}
Then, the condition (\ref{ins2}) for avoiding the 
Laplacian instability of scalar perturbations reads
\begin{equation}
\frac{d}{dt} \left( \frac{4L_\s^2}{{\cal W}} \right)
+\frac{4HL_\s^2}{{\cal W}}-\mathcal{E}>0\,,
\end{equation}
where $L_\s$, $\mathcal{E}$, and ${\cal W}$ are given by 
Eqs.~(\ref{Ls}), (\ref{calE}), and (\ref{Wre}) respectively.

As an example, let us consider BD theory described by 
the functions (\ref{BDaction}). 
Since $L_\s=\mathcal{E}=G_4=M_{\rm pl}\phi/2$ 
in this case, the conditions (\ref{Ls}) and (\ref{calE}) 
are satisfied for 
\begin{equation}
\phi>0\,,
\label{BDCon1}
\end{equation}
with the tensor propagation speed square $c_t^2=1$.
Since ${\cal W}=M_{\rm pl} (\dot{\phi}+2H \phi)$ and 
$w=-3M_{\rm pl}(6H^2 \phi^2-\omega_{\rm BD}\dot{\phi}^2
+6H \phi \dot{\phi})/(2\phi)$, the quantity (\ref{Qsho}) reads
\begin{equation}
Q_s=\frac{(3+2\omega_{\rm BD})M_{\rm pl}\phi
\dot{\phi}^2}{(\dot{\phi}+2H\phi)^2}\,.
\label{BDCon2}
\end{equation}
On using the condition (\ref{BDCon1}), we find that the 
scalar ghost is absent for 
\begin{equation}
\omega_{\rm BD}>-3/2\,.
\label{BDCon3}
\end{equation}
The quantity ${\cal M}$ can be expressed as
\begin{equation}
\M=-\frac{M_{\rm pl}^2 \phi^2}{{\cal F}}\,,
\end{equation}
where we have used the fact that the term ${\cal F}$ in Eq.~(\ref{Fdef})
is given by ${\cal F}=-M_{\rm pl}(\dot{\phi}+2H \phi)$.
{}From the background equation (\ref{back3}), it follows that 
\begin{equation}
\dot{\cal F}=-L_N=-M_{\rm pl} \dot{\phi}
(3H\phi-\omega_{\rm BD} \dot{\phi})/\phi\,.
\end{equation}
Then, the condition (\ref{ins2}) for avoiding the Laplacian instability 
of scalar perturbations translates to 
\begin{equation}
\dot{{\cal M}}+H{\cal M}-{\cal E}
=\frac{(3+2\omega_{\rm BD})M_{\rm pl}\phi
\dot{\phi}^2}{2(\dot{\phi}+2H\phi)^2}>0\,,
\end{equation}
which is satisfied under (\ref{BDCon1}) and (\ref{BDCon3}).
In fact, from Eq.~(\ref{csdef}), the scalar propagation speed square 
$c_s^2$ is equivalent to 1 in BD theory.

\subsection{Primordial power spectra in k-inflation}

Let us consider a non-canonical scalar-field theory
described by the Lagrangian (\ref{klag}).
This theory can be expressed in terms of the ADM
variables as Eq.~(\ref{multilag2}).
Since $L_{\s}={\cal E}=G_4=M_{\rm pl}^2/2$, $Q_t=M_{\rm pl}^2/4$ and 
$c_t^2=1$, the tensor mode is not plagued by any ghosts and 
Laplacian instabilities. 
{}From Eq.~(\ref{tensorpower}), the tensor 
power spectrum is given by 
\begin{equation}
{\cal P}_h=\frac{2H^2}{\pi^2 M_{\rm pl}^2}\,,
\label{Ph}
\end{equation}
which depends only on $H$.
Therefore, if the amplitude of primordial gravitational waves 
is measured, the energy scale of inflation can be explicitly known.

We also have the relations ${\cal W}=2HM_{\rm pl}^2$, 
$w=-9H^2 M_{\rm pl}^2+3X(P_{X}+2X P_{XX})$, and
\begin{equation}
Q_s=-\frac{\dot{\phi}^2(P_{X}+2X P_{XX})}{H^2}\,,
\label{kinf1}
\end{equation}
so the scalar ghost is absent for $P_{X}+2X P_{XX}<0$.
Since ${\cal F}=-2M_{\rm pl}^2 \dot{H}$ and 
$L_N=2\dot{\phi}^2P_{X}$, the background equation 
of motion (\ref{back3}) gives 
$M_{\rm pl}^2 \dot{H}=\dot{\phi}^2 P_{X}$.
Taking the time derivative of the quantity 
${\cal M}=M_{\rm pl}^2/(2H)$, 
it follows that 
\begin{equation}
\dot{{\cal M}}+H{\cal M}-{\cal E}
=-\frac{M_{\rm pl}^2 \dot{H}}{2H^2}
=-\frac{\dot{\phi}^2P_{X}}{2H^2}\,.
\label{kinf2}
\end{equation}
To avoid the instability of scalar perturbations, 
we require that $P_{X}<0$.
Substituting Eqs.~(\ref{kinf1}) and (\ref{kinf2}) 
into Eq.~(\ref{csdef}), we obtain 
\begin{equation}
c_s^2=\frac{P_{X}}{P_{X}+2XP_{XX}}\,.
\label{cskinf}
\end{equation}
In standard slow-roll inflation driven by the potential energy 
$V(\phi)$ of a canonical scalar field ($P=-X/2-V(\phi)$), 
$c_s^2$ is equivalent to 1. 
If the Lagrangian $P$ contains a non-linear term in $X$,
the scalar propagation speed is generally different from 1.

{}From Eqs.~(\ref{kinf1}) and (\ref{cskinf}), we find that 
the slow-roll parameter $\epsilon=-\dot{H}/H^2$ 
is related to $Q_s$ and $c_s^2$, as
\begin{equation}
\epsilon=\frac{Q_s c_s^2}{M_{\rm pl}^2}\,.
\end{equation}
Then, the scalar power spectrum (\ref{scalarpower}) reads
\begin{equation}
{\cal P}_{\zeta}=\frac{H^2}{8\pi^2M_{\rm pl}^2 \epsilon c_s}\,.
\label{Pzeta2}
\end{equation}
{}From Eqs.~(\ref{Ph}) and (\ref{Pzeta2}), 
the tensor-to-scalar ratio is given by \cite{kinfper}
\begin{equation}
r=16c_s \epsilon\,.
\end{equation}
Since $\epsilon \ll 1$ during inflation, 
it follows that $r \ll 1$ for $c_s \leq 1$. 

\section{Horndeski theory in the language of EFT}
\label{effsec}

In this section, we relate the variables introduced in Sec.~\ref{actionsec} 
with those employed in the EFT language of Refs.~\cite{Cheung,Gubi,Bloom}. 
The action expanded up to quadratic order in the perturbations 
can be written in the following form 
\begin{eqnarray}
S &=& \int d^4 x\sqrt{-g} \biggl[ \frac{M_*^2}{2}fR
-\Lambda-c\,g^{00} 
+\frac{M_2^4}{2} (\delta g^{00})^2
-\frac{\bar{m}_1^3}{2} \delta K \delta g^{00}
-\frac{\bar{M}_2^2}{2} \delta K^2 
\nonumber \\ & &
-\frac{\bar{M}_3^2}{2} \delta K^{\mu}_{\nu} \delta K^{\nu}_{\mu}
+\frac{\mu_1^2}{2}{\cal R} \delta g^{00}
+\frac{\bar{m}_5}{2}{\cal R} \delta K
+\frac{\lambda_1}{2}{\cal R}^2 
+\frac{\lambda_2}{2} {\cal R}^{\mu}_{\nu} 
{\cal R}^{\nu}_{\mu} \biggr],
\label{action2}
\end{eqnarray}
where $g^{00}=-1/N^2$, $M_*$ is a constant, and 
other coefficients such as $f, \Lambda, c, M_2^4$ 
depend on time.
We note that the four-dimensional Ricci scalar $R$ can 
be written in terms of the three-dimensional quantities as 
Eq.~(\ref{Rre}). After integration by parts,
the first term in Eq.~(\ref{action2}) reads
\begin{equation}
\frac{M_*^2}{2}fR
=\frac{M_*^2}{2} \left( f{\cal R}
+f {\cal S}-fK^2-2\dot{f}\frac{K}{N} \right)\,.
\label{fR}
\end{equation}

Now we substitute ${\cal R}=\delta_1 {\cal R}+\delta_2 {\cal R}$, 
$K=3H^2+\delta K$, and 
${\cal S}=3H^2+2H\delta K+ \delta K^{\mu}_{\nu} \delta K^{\nu}_{\mu}$
into Eq.~(\ref{fR}) and then expand the action (\ref{action2}) up to 
quadratic order in the perturbations. 
In doing so, we use the similar property to Eq.~(\ref{delK}), i.e., 
$\int d^4x \sqrt{-g}\,\beta(t) \delta K=
\int d^4 x \sqrt{-g} (-\dot{\beta}-3H\beta+\dot{\beta}\delta N
-\dot{\beta}\delta N^2)$, where $\beta(t)$ is an arbitrary function 
in terms of $t$.
Then, the resulting Lagrangian reads
\begin{eqnarray}
L &=&
M_*^2 (\ddot{f}+2H \dot{f}+2\dot{H}f+3H^2 f)-\Lambda+c 
\nonumber \\
&& 
+[M_*^2 (-\ddot{f}+H \dot{f}-2\dot{H}f)-2c]
\delta N+\frac{M_*^2}{2}f \delta_1 {\cal R} \nonumber \\
&& 
+\left[M_*^2 (\ddot{f}-H \dot{f}+2\dot{H}f)+3c+2M_2^4 \right]
\delta N^2-\left( \frac{M_*^2}{2}f+\frac{\bar{M}_2^2}{2} 
\right)\delta K^2 \nonumber \\
&& 
+(M_*^2 \dot{f}-\bar{m}_1^3)\delta K \delta N+
\frac{\bar{m}_5}{2}\delta K \delta_1 {\cal R}
+\mu_1^2 \delta N \delta_1 {\cal R}
+\frac{M_*^2}{2}f \delta_2 {\cal R}  \nonumber \\
&& 
+
\left( \frac{M_*^2}{2}f-\frac{\bar{M}_3^2}{2} \right)
\delta K^{\mu}_{\nu} \delta K^{\nu}_{\mu}
+\frac{\lambda_1}{2}{\cal R}^2
+\frac{\lambda_2}{2}
\delta {\cal R}^{\mu}_{\nu} \delta {\cal R}^{\nu}_{\mu}\,.
\label{Leff}
\end{eqnarray}
Comparing the terms up to the second line of Eq.~(\ref{Leff}) 
with those in Eq.~(\ref{lagfirst}), it follows that 
\begin{eqnarray}
& & M_*^2 (\ddot{f}+2H \dot{f}+2\dot{H}f+3H^2 f)-\Lambda+c 
=\bar{L}- \dot{{\cal F}}-3H{\cal F}\,,\label{bac1}\\
& & M_*^2 (-\ddot{f}+H \dot{f}-2\dot{H}f)-2c=
\dot{\cal F}+L_N\,,\label{bac2}\\
& & f=\frac{2}{M_*^2}{\cal E}
=\frac{1}{M_*^2} \left( 2L_{\cal R}+\dot{L_\U}
+3H L_{\U} \right)\,.
\label{bac3}
\end{eqnarray}
{}From Eqs.~(\ref{back2}) and (\ref{back3}), 
the r.h.s. of Eq.~(\ref{bac1}) and (\ref{bac2}) vanish 
in the absence of matter. 
The background equations of motion are characterized by 
the three parameters $f$, $\Lambda$, and $c$.
Comparing the second-order terms in Eq.~(\ref{Leff}) with 
those in Eq.~(\ref{lagex}), 
we obtain the following relations
\begin{eqnarray}
& & M_2^4=\frac14 (2L_N+L_{NN}-2c)\,,\qquad
\bar{m}_1^3=2\dot{{\cal E}}-L_{KN}-2HL_{\s N}\,,\nonumber \\
& &\bar{M}_2^2=-2{\cal E}-L_{KK}-4HL_{\s K}-4H^2 L_{\s \s}\,,\qquad
\bar{M}_3^2=2{\cal E}-2L_{\s}\,, \nonumber \\
& & \mu_1^2=L_{N{\cal R}}-\frac12 \dot{L}_{\U}+H L_{N \U}\,,
\nonumber \\
& &\bar{m}_5=2L_{K{\cal R}}+4H L_{\s{\cal R}}+L_{\U}+
2H L_{K \U}+4H^2 L_{\s \U}\,,\nonumber \\
& &\lambda_1=L_{{\cal R}{\cal R}}+2H L_{{\cal R} \U}
+H^2 L_{\U \U}\,,\qquad
\lambda_2=2L_{\Z}\,,
\label{corres}
\end{eqnarray}
where we have used Eq.~(\ref{bac2}) to derive $M_2^4$.
In Horndeski theory, the r.h.s. of Eq.~(\ref{corres}) 
can be evaluated by taking partial derivatives 
of the Lagrangian (\ref{Ltotal}) in terms of 
the scalar variables.

The conditions (\ref{elicon1})-(\ref{elicon3}) reduce, 
respectively, to
\begin{equation}
\bar{M}_2^2+\bar{M}_3^2=0\,,\qquad
\bar{m}_5=0\,,\qquad
8\lambda_1+3\lambda_2=0\,,
\end{equation}
under which the spatial derivatives higher than second 
order are absent.
On using these conditions, the Lagrangian (\ref{action2}) can 
be expressed as 
\begin{eqnarray}
S &=& \int d^4 x\sqrt{-g} \biggl[ \frac{M_*^2}{2}fR
-\Lambda-c\,g^{00} 
+\frac{M_2^4}{2} (\delta g^{00})^2
-\frac{\bar{m}_1^3}{2} \delta K \delta g^{00}
\nonumber \\ & &
~~~~~~~~~~~~~~~~~~
-m_4^2 \left( \delta K^2 -\delta K^{\mu}_{\nu} 
\delta K^{\nu}_{\mu} \right) 
+\frac{\mu_1^2}{2}{\cal R} \delta g^{00}\biggr]\,,
\label{action3}
\end{eqnarray}
where 
\begin{equation}
m_4^2 \equiv \frac14 \left( \bar{M}_2^2 
-\bar{M}_3^2 \right)
=\frac14 \left( -4{\cal E}+2L_{\s}-L_{KK}
-4H L_{\s K}-4H^2 L_{\s \s} \right)\,.
\end{equation}
The terms containing ${\cal R}^2=16(\partial^2 \zeta)^2/a^4$ 
and ${\cal R}_{ij}{\cal R}^{ij}=[5(\partial^2 \zeta)^2
+(\partial_i \partial_j \zeta)^2]/a^4$ are absent in Eq.~(\ref{action3})
because they only involve spatial derivatives of $\zeta$ higher than 
second order.

In Horndeski theory described by the action (\ref{Ltotal}), 
the coefficients in the action (\ref{action3}) can be computed
by using Eqs.~(\ref{bac1})-(\ref{corres}).
They are given by 
\begin{eqnarray}
\hspace{-0.3cm}
M_*^2 f &=& 2G_4-G_{5\phi}\dot{\phi}^2
+2G_{5X}\dot{\phi}^2\ddot{\phi}\,,
\label{effq1}\\
\hspace{-0.3cm}
\Lambda &=& 
XG_{2X}-G_2+\dot{\phi}^2(\ddot{\phi}+3H \dot{\phi})G_{3X}
+\dot{\cal F}_4/2+3H\dot{X}G_{4X}
-18H^2G_{4X}\dot{\phi}^2\nonumber\\
&&
+6HG_{4\phi X}\dot{\phi}^3
+12H^2G_{4XX} \dot{\phi}^4
+\dot{\cal F}_5/2+3M_*^2H^2 f_5 
+3M_*^2 H\dot{f}_5/2 \nonumber\\
&&
-6H^2 G_{5\phi} \dot{\phi}^2
-7H^3G_{5X} \dot{\phi}^3+3H^2G_{5\phi X} \dot{\phi}^4
+2H^3G_{5XX} \dot{\phi}^5\,,\label{effq2} \\
\hspace{-0.3cm}
c &=&XG_{2X}+\dot{\phi}^2(-\ddot{\phi}+3H \dot{\phi})G_{3X}
+\dot{\phi}^2G_{3\phi}
-\dot{\cal F}_4/2+3H \dot{X} G_{4X}\nonumber\\ &&
-6H^2G_{4X}\dot{\phi}^2
+6HG_{4\phi X} \dot{\phi}^3 
+12H^2G_{4XX} \dot{\phi}^4
-\dot{\cal F}_5/2+3M_*^2 H\dot{f}_5/2
\nonumber\\ &&
-3H^2 G_{5\phi} \dot{\phi}^2
-3H^3G_{5X} \dot{\phi}^3+3H^2 G_{5\phi X} \dot{\phi}^4
+2H^3 G_{5XX} \dot{\phi}^5\,,
\label{effq3} \\
\hspace{-0.3cm}
M_2^4 &=& X^2 G_{2XX}+(\ddot{\phi}+3H \dot{\phi})
G_{3X}\dot{\phi}^2/2
-3HG_{3XX} \dot{\phi}^5-G_{3\phi X} \dot{\phi}^4/2
\nonumber \\
& &
+\dot{\cal F}_4/4
-3H \dot{X}G_{4X}/2
+6HG_{4\phi X} \dot{\phi}^3
+18H^2 G_{4XX}\dot{\phi}^4-6HG_{4\phi XX}\dot{\phi}^5
\nonumber \\ 
& &
-12H^2 G_{4XXX} \dot{\phi}^6
+\dot{\cal F}_5/4-3M_*^2H\dot{f}_5/4-3H^3 G_{5X} \dot{\phi}^3/2
\nonumber \\ & &
+6H^2 G_{5\phi X} \dot{\phi}^4
+6H^3 G_{5XX}\dot{\phi}^5
-3H^2 G_{5\phi XX}\dot{\phi}^6
-2H^3G_{5XXX}\dot{\phi}^7, \\
\hspace{-0.3cm}
\bar{m}_1^3 &=& 2G_{3X}\dot{\phi}^3+2\dot{X}G_{4X}
-8HG_{4X}\dot{\phi}^2+4G_{4\phi X}\dot{\phi}^3
+16HG_{4XX}\dot{\phi}^4\,,
\nonumber \\
& & +M_*^2 \dot{f}_5-4HG_{5\phi}\dot{\phi}^2
-6H^2G_{5X} \dot{\phi}^3
+4HG_{5\phi X} \dot{\phi}^4
+4H^2G_{5XX}\dot{\phi}^5,\\
\hspace{-0.3cm}
m_4^2 &=& \mu_1^2=2G_{4X}\dot{\phi}^2
+G_{5\phi}\dot{\phi}^2
+HG_{5X}\dot{\phi}^3-G_{5X}\dot{\phi}^2 \ddot{\phi} \,,
\label{corre9}
\end{eqnarray}
where 
\begin{eqnarray}
{\cal F}_4 &=&
2\dot{X}G_{4X}-8HG_{4X}\dot{\phi}^2\,,\\
{\cal F}_5 &=& 
2M_*^2 H f_5+M_*^2\dot{f}_5-2HG_{5\phi} \dot{\phi}^2
-2H^2G_{5X} \dot{\phi}^3\,,\\
M_*^2 f_5 &=&
-G_{5\phi}\dot{\phi}^2+2G_{5X}\dot{\phi}^2 \ddot{\phi}\,.
\end{eqnarray}

We stress that Horndeski theory satisfies the additional
relation $m_4^2= \mu_1^2$.
The time and spatial derivatives for the theory (\ref{action3})
are kept up to second order for {\it linear} cosmological perturbations. 
If $m_4^2 \neq \mu_1^2$, then higher-order 
spatial derivatives should appear beyond linear order. 
For the computation of primordial non-Gaussianities of 
curvature perturbations generated during inflation, we 
need to expand the action (\ref{action0}) higher than quadratic order.  
In such cases, the presence of higher-order spatial derivatives 
can modify the shape of non-Gaussianities \cite{ng1,Cremi10} 
relative to that derived for Horndeski theory \cite{XGao,Tsujinon2,Hornshape}.
  
\section{Application to dark energy}
\label{darksec}

In this section, we study the dynamics of dark energy based on 
Horndeski theory in the presence of matter 
(cold dark matter, baryons, photons e.t.c.).
The action in such a theory is given by 
\begin{equation}
S=\int d^4x \sqrt{-g} \,\sum_{i=2}^{5} L_i+
\int d^4 x\,L_m\,,
\label{Hoac}
\end{equation}
where $L_{2,3,4,5}$ are given by Eqs.~(\ref{eachlag2})-(\ref{L5lag}) 
and $L_m$ is the matter Lagrangian of a barotropic perfect fluid.
The scalar degree of freedom is responsible for 
the late-time cosmic acceleration.
We assume that matter does not have a direct coupling to $\phi$.

\subsection{Background equations of motion}

On the flat FLRW background, the energy-momentum tensor 
of the barotropic perfect fluid 
is given by $T^0_0=-\rho_m$ and $T^i_j=P_m\delta^i_j$, 
where $\rho_m$ is the energy density and $P_m$ is the pressure.
This satisfies the continuity equation $T^{\mu}_{0;\mu}=0$, i.e., 
\begin{equation}
\dot{\rho}_m+3H(\rho_m+P_m)=0\,.
\label{continuity}
\end{equation}
In the presence of matter, the background equations of motion 
(\ref{back1}) and (\ref{back3}) are modified to 
\begin{eqnarray}
& &
\bar{L}+L_N-3H {\cal F}=\rho_m\,,
\label{backmo1} \\
& &
\dot{\cal F}+L_N=\rho_m+P_m\,.
\label{backmo2} 
\end{eqnarray}
Substituting Eqs.~(\ref{backmo1})-(\ref{backmo2}) into 
Eqs.~(\ref{bac1})-(\ref{bac2}), we obtain 
\begin{eqnarray}
\Lambda+c &=& 3M_*^2 (fH^2+\dot{f}H)-\rho_m\,,\label{lam1}\\
\Lambda-c &=& M_*^2 (2f\dot{H}+3fH^2+2\dot{f}H+\ddot{f})+P_m\,.
\label{lam2}
\end{eqnarray}
In Horndeski theory, the functions $f$, $\Lambda$, $c$ are 
given, respectively, by Eqs.~(\ref{effq1}), (\ref{effq2}), and (\ref{effq3}).
Among the four functions $G_{2,3,4,5}$, the three combinations
of them (i.e., $f,\Lambda,c$) determine the cosmological dynamics.

Taking the time derivative of Eq.~(\ref{lam1}) and using 
Eqs.~(\ref{continuity}) and (\ref{lam2}), we obtain 
\begin{equation}
\dot{\Lambda}+\dot{c}+6Hc
=3M_*^2 \dot{f} (2H^2 +\dot{H})\,.
\label{dotc}
\end{equation}
The background equations of motion (\ref{lam1}) and 
(\ref{lam2}) can be expressed as
\begin{eqnarray}
& & 3M_{\rm pl}^2 H^2=\rho_{\rm DE}+\rho_m\,,\label{be1}\\
& & M_{\rm pl}^2 (2\dot{H}+3H^2)=-P_{\rm DE}-P_m\,,
\end{eqnarray}
where
\begin{eqnarray}
\rho_{\rm DE} &=& c+\Lambda+3H^2 (M_{\rm pl}^2-M_*^2 f)
-3M_*^2 \dot{f}H\,,\\
P_{\rm DE} &=& c-\Lambda-(2\dot{H}+3H^2)
(M_{\rm pl}^2-M_*^2 f)+M_*^2 (2H \dot{f}+\ddot{f})\,.
\end{eqnarray}
On using Eq.~(\ref{dotc}), we find that the ``dark'' component 
satisfies the standard continuity equation 
\begin{eqnarray}
\dot{\rho}_{\rm DE}+3H (\rho_{\rm DE}+P_{\rm DE})=0\,.
\end{eqnarray}
Then, we can define the equation of state of dark energy, as 
\begin{equation}
w_{\rm DE}=\frac{P_{\rm DE}}{\rho_{\rm DE}}
=-1+\frac{2c-2\dot{H} (M_{\rm pl}^2-M_*^2 f)
-M_*^2 (H \dot{f}-\ddot{f})}
{c+\Lambda+3H^2 (M_{\rm pl}^2-M_*^2 f)
-3M_*^2 \dot{f}H}\,.
\end{equation}

For quintessence described by the Lagrangian $G_2=P(\phi,X)$, 
$G_3=0$, $G_4=M_{\rm pl}^2/2$, and $G_5=0$, we have 
$M_*^2 f=M_{\rm pl}^2$, $\Lambda=V(\phi)$, and 
$c=\dot{\phi}^2/2$. Since 
$w_{\rm DE}=[\dot{\phi}^2/2-V(\phi)]/[\dot{\phi}^2/2+V(\phi)]$ 
in this case, it follows that $w_{\rm DE}>-1$.
For a non-canonical scalar field with the Lagrangian 
(\ref{klag}) we have $w_{\rm DE}<-1$ 
for $P_{X}>0$, but the scalar ghost is present.
For the theories in which the quantity $f$ varies 
in time (i.e., $G_4$ or $G_5$ varies), 
it is possible to realize $w_{\rm DE}<-1$ under the condition 
\begin{equation}
2c-2\dot{H} (M_{\rm pl}^2-M_*^2 f)
-M_*^2 (H \dot{f}-\ddot{f})<0\,,
\end{equation}
where we have assumed $\rho_{\rm DE}>0$.
In $f(R)$ gravity \cite{fRviable1,fRviable2,fRviable3,fRviable4,fRviable5} 
and Galileons \cite{Galicosmo}, 
the dark energy equation of state can be smaller than $-1$, 
while avoiding the appearance of ghosts.

\subsection{Matter density perturbations and effective gravitational couplings}

Let us proceed to discuss the equations of motion for linear cosmological perturbations. 
The discussion in Sec.~\ref{actionsec} is based on unitary gauge, but for the study 
of dark energy, the Newtonian gauge is commonly used. 
The general metric in the presence of scalar perturbations $\Psi$, $\psi$, 
$\Phi$, and $E$ can be written as 
\begin{equation}
ds^2=-(1+2\Psi) dt^2+2\psi_{|i} dx^i dt+a^2(t) 
\left[ (1+2\Phi)\delta_{ij}+\partial_{ij}E \right]dx^i dx^j\,.
\label{genemet}
\end{equation}
The Newtonian gauge corresponds to $\psi=0$ and $E=0$.

Since the Horndeski action is equivalent to the EFT action (\ref{action3}) 
in unitary gauge with $m_4=\mu_1^2$ (up to second order), 
it is possible to derive the perturbation equations in 
general gauge by reintroducing 
the scalar perturbation $\delta \phi$ via the Stueckelberg 
trick \cite{Cremi,Cheung,Bloom2,Piazza}. 
The quantities appearing in the action (\ref{action3}) transform 
under the time coordinate change $t \to t+\delta \phi(t,{\bm x})$, 
e.g., $\delta K_{ij} \to \delta K_{ij}-\dot{H} \delta \phi h_{ij}-\partial_i \partial_j \delta \phi$, 
${}^{(3)}R_{ij} \to {}^{(3)}R_{ij}+H(\partial_i \partial_j \delta \phi+\delta_{ij} \partial^2 \delta \phi)$.
This transformation allows one to write the action (\ref{action0}) up to 
quadratic order in the perturbations for the general metric (\ref{genemet}).
Varying the resulting action $S$ with respect to $\Psi$, $\psi$, $\Phi$, 
$E$, $\delta \phi$ and finally setting $\psi=0=E$, we can derive the perturbation 
equations in the Newtonian gauge. This is the approach taken 
in Ref.~\cite{Piazza}.

As performed in Ref.~\cite{Koba}, the perturbation equations can 
be also derived by directly expanding the Horndeski action (\ref{Hoac}) 
for the metric (\ref{genemet}). 
In the following we assume that the matter Lagrangian $L_m$ 
is described by a barotropic perfect fluid of non-relativistic matter
with the energy-momentum tensor 
\begin{equation}
T^0_0=-(\rho_m+\delta \rho_m)\,,\qquad
T^0_i=-\rho_m \partial_i v_m\,,\qquad
T^i_j=0\,.
\end{equation}
Since there is no direct coupling between matter and the field $\phi$, 
the perturbed energy-momentum tensor obeys the continuity equation 
\begin{equation}
{\delta T^{\mu \nu}}_{;\mu}=0\,.
\label{delTcon}
\end{equation}
{}From the $\nu=0$ and $\nu=i$ components of Eq.~(\ref{delTcon}), 
we obtain the following equations in Fourier space respectively,
\begin{eqnarray}
& &
\dot{\delta \rho_m}+3H \delta \rho_m+
3\rho_m \dot{\Phi}+\frac{k^2}{a^2} \rho_m v_m=0\,,
\label{mattereq1}\\
& &
\dot{v}_m=\Psi\,,
\label{mattereq2}
\end{eqnarray}
where $k$ is a comoving wavenumber.
We introduce the gauge-invariant density contrast 
\begin{equation}
\delta_m \equiv \frac{\delta \rho_m}{\rho_m}
+3H v_m\,.
\label{delmdef}
\end{equation}
Taking the time derivative of (\ref{mattereq1}) and 
using Eq.~(\ref{mattereq2}), the density contrast satisfies
\begin{equation}
\ddot{\delta}_m+2H \dot{\delta}_m
+\frac{k^2}{a^2}\Psi=3\ddot{Q}+6H \dot{Q}\,,
\label{mattereq}
\end{equation}
where $Q \equiv Hv_m-\Phi$.

Expanding the action (\ref{Hoac}) for the metric (\ref{genemet}) 
up to quadratic order in the perturbations, varying the resulting action 
with respect to $E$, $\Psi$, $\delta \phi$, and finally setting $\psi=E=0$, 
we obtain the following perturbation equations respectively:
\begin{eqnarray}
\hspace{-0.7cm} & &
B_6 \Phi+B_7 \delta \phi+B_8 \Psi=0\,,
\label{ma3} \\
\hspace{-0.7cm} & &
A_{1}\dot{\Phi}+A_{2}\dot{\delta\phi}-\rho_{m}\Psi
+B_8\frac{k^{2}}{a^{2}}\Phi+A_{4}\Psi
+\left(A_{6}\frac{k^{2}}{a^{2}}-\mu\right)\delta\phi-\delta \rho_{m}=0\,,
\label{ma1}\\
\hspace{-0.7cm}& & 
D_{1}\ddot{\Phi}+D_{2}\ddot{\delta\phi}+D_{3}\dot{\Phi}
+D_{4}\dot{\delta\phi}+D_{5}\dot{\Psi}+\left(B_{7}\frac{k^{2}}{a^{2}}
+D_{8}\right)\Phi \nonumber \\
\hspace{-0.7cm}& &
+\left(D_{9}\frac{k^{2}}{a^{2}}-M^{2}\right)\delta\phi
+\left(A_6\frac{k^{2}}{a^{2}}+D_{11}\right)\Psi=0\,,
\label{ma2} 
\end{eqnarray}
where 
\begin{eqnarray}
B_6 &=& 4{\cal E}
=4G_4+2XG_{5\phi}-4XG_{5X} \ddot{\phi}\,,\label{B6}\\
B_7 &=&
\frac{4}{\dot{\phi}} \left[ \dot{L_{\cal S}}+H (L_{\cal S}-{\cal E}) 
\right]\,, \nonumber \\
&=& 8G_{4X}H \dot{\phi}+8(G_{4X}+2X G_{4XX}) \ddot{\phi}
+4G_{4\phi}-8X G_{4\phi X} \nonumber \\
& &+4(G_{5\phi}+XG_{5\phi X}) \ddot{\phi}
+4H \left[ 2(G_{5X}+XG_{5XX})\ddot{\phi}
+G_{5\phi}-XG_{5\phi X} \right]\dot{\phi}\nonumber \\
& &
-2XG_{5\phi \phi}-4(H^2+\dot{H})XG_{5X}\,,
\\
B_8&=& 4L_{\cal S}=4G_4-8XG_{4X}
-4H \dot{\phi}XG_{5X}-2X G_{5\phi}\,.
\label{B8}
\end{eqnarray}
Explicit forms of the time-dependent coefficients 
$A_i$ and $D_i$ as well as other perturbations equations 
(derived by the variations of $\Phi$ and $\psi$)
are given in Ref.~\cite{Koba}.
The definition of the term $\mu$ in Eq.~(\ref{ma1}) 
is $\mu={\cal H}_{\phi}$, where ${\cal H} \equiv -(\bar{L}+L_N-3H{\cal F})$.
The term $M$ in Eq.~(\ref{ma2}) is defined by 
\begin{equation}
M^2 \equiv \left[ \dot{\mu}+3H (\mu +\nu) \right]/\dot{\phi}\,,
\end{equation}
where $\nu={\cal P}_{\phi}$ with 
${\cal P} \equiv \bar{L}-\dot{\cal F}-3H {\cal F}$.
The mass square $M^2$ involves the second derivative 
of $-G_2$ with respect to $\phi$ \cite{Koba}. 
For a canonical field with the potential $V(\phi)$,
this means that the second derivative $V_{\phi \phi}$ is present 
in the expression of $M^2$. 
For dark energy models in which the so-called chameleon 
mechanism \cite{chameleon} works to suppress the fifth force 
mediated by the field $\phi$, the models are designed to have a large 
mass $M$ in the region of high 
density \cite{fRviable1,fRviable2,fRviable3,fRviable4,fRviable5,chamesca}.
In the low-energy regime where the late-time cosmic acceleration 
comes into play, the mass $M$ should be as small as $H_0$.

The perturbations related to the observations of large-scale 
structures and weak lensing have been
deep inside the Hubble radius in the low-redshift regime.
In the following we use the quasi-static approximation on 
sub-horizon scales, under which the dominant contributions
to Eqs.~(\ref{ma1}) and (\ref{ma2}) are those involving 
the terms $k^2/a^2$, $\delta \rho_m$, and 
$M^2$ \cite{quasi,Bloom2}. 
In doing so, we neglect the contribution of the oscillating 
term of the field perturbation $\delta \phi$ relative to 
the one induced from the matter perturbation $\delta \rho_m$. 
Under this approximation scheme, the variations of the 
gravitational potentials $\Phi$ and $\Psi$ are small
such that $|\dot{\Phi}| < |H \Phi|$ and 
$|\dot{\Psi}| < |H \Psi|$.
Then, Eqs.~(\ref{ma1}) and (\ref{ma2}) read
\begin{eqnarray}
& &
B_8\frac{k^{2}}{a^{2}}\Phi
+A_{6}\frac{k^{2}}{a^{2}}\delta\phi
-\delta \rho_{m} \simeq 0\,,\label{ma1d}\\
& &B_{7}\frac{k^{2}}{a^{2}}\Phi+
\left( D_9 \frac{k^2}{a^2}-M^2 \right) \delta \phi
+A_6 \frac{k^2}{a^2}\Psi \simeq 0\,,
\label{ma2d}
\end{eqnarray}
where 
\begin{eqnarray}
A_6 &=&
2XG_{3X}+8H (G_{{4X}}+2XG_{4XX})
\dot{\phi}+2G_{{4\phi}}+4XG_{{4\phi X}}\nonumber \\
& &+4H\left(G_{{5\phi}}+XG_{{5\phi X}}\right)\dot{\phi}
-2{H}^{2}X\left(3G_{{5X}}+2XG_{{5XX}}\right)\,,\\
D_{9} & = & 2G_{2X}-4\left(G_{{3X}}+XG_{{3XX}}\right)\ddot{\phi}
-8HG_{3X}\dot{\phi}-2G_{{3\phi}}+2XG_{3\phi X} \nonumber \\
&  & +[-16H(3\, G_{{4XX}}+2XG_{{4XXX}})\ddot{\phi}
-8H(3G_{4\phi X}-2XG_{{4\phi XX}})]\dot{\phi}
\nonumber \\
&  &
-4(3 G_{{4\phi X}}+2XG_{{4\phi XX}})\ddot{\phi}
+40{H}^{2}XG_{{4 XX}}+4XG_{{4\phi\phi X}} \nonumber \\
&  & 
+8\dot{H}(G_{4X}+2XG_{{4XX}})+12{H}^{2}G_{4X}
+\{-8H(2G_{{5\phi X}}+XG_{{5\phi{\it XX}}})\ddot{\phi}
\nonumber \\
&  & 
+8H(H^{2}+\dot{H})(G_{{5X}}+XG_{{5{\it XX}}})
+4HXG_{5\phi \phi X}\}\dot{\phi}
-4H^{2}X^{2}G_{{5\phi{\it XX}}}\nonumber \\
 &  & +4H^{2}(G_{{5X}}+5XG_{{5{\it XX}}}
+2{X}^{2}G_{{5{\it XXX}}})\ddot{\phi}
+2(3H^{2}+2\dot{H})G_{5\phi}
\nonumber \\
&  & 
+4\dot{H}XG_{{5\phi X}}+10{H}^{2}XG_{{5\phi X}}\,.
\end{eqnarray}
Solving Eqs.~(\ref{ma3}), (\ref{ma1d}), and (\ref{ma2d}) 
for $\Psi$ and $\Phi$, it follows that 
\begin{eqnarray}
& &
\frac{k^2}{a^2}\Psi \simeq 
-{\frac{(B_{6}D_{9}-B_{7}^{2})\,{(k/a)}^{2}-B_{6}M^{2}}
{(A_{6}^{2}B_{6}+B_{8}^{2}D_{9}-2A_{6}B_{7}B_{8})\,
{(k/a)}^{2}-B_{8}^{2}M^{2}}}\delta \rho_m\,,\label{Psiso}\\
& &
\frac{k^2}{a^2}\Phi \simeq 
-{\frac{(A_{6}B_{7}-B_{8}D_{9})\,{(k/a)}^{2}+B_{8}M^{2}}
{(A_{6}^{2}B_{6}+B_{8}^{2}D_{9}-2A_{6}B_{7}B_{8})\,
{(k/a)}^{2}-B_{8}^{2}M^{2}}}\delta \rho_m\,.\label{Phiso}
\end{eqnarray}

From Eq.~(\ref{mattereq1}), we find that the term $Hv_m$  
is at most of the order of $(aH/k)^2 \delta \rho_m/\rho_m$. 
For the modes deep inside the Hubble radius ($k \gg aH$),
we then have $\delta_m \simeq \delta \rho_m/\rho_m$ 
in Eq.~(\ref{delmdef}). Under the quasi-static approximation 
on sub-horizon scales, the r.h.s. of Eq.~(\ref{mattereq}) is
negligible relative to the l.h.s. of it.  
On using Eq.~(\ref{Psiso}), the linear 
matter perturbation obeys 
\begin{equation}
\ddot{\delta}_m+2H \dot{\delta}_m-4\pi G_{\rm eff} 
\rho_m \delta_m \simeq 0\,,
\label{mattereqf}
\end{equation}
where 
\begin{equation}
G_{\rm eff}={\frac{2\Mpl^{2}[(B_{6}D_{9}-B_{7}^{2})\,{(k/a)}^{2}
-B_{6}M^{2}]}{(A_{6}^{2}B_{6}+B_{8}^{2}D_{9}-2A_{6}B_{7}B_{8})\,
{(k/a)}^{2}-B_{8}^{2}M^{2}}}G\,.
\label{Geff}
\end{equation}
Note that $G$ is the bare gravitational constant related with the reduced
Planck mass $M_{{\rm pl}}$ via the relation $8\pi G=M_{{\rm pl}}^{-2}$. 
Since the effective gravitational coupling $G_{\rm eff}$ 
is different depending on gravitational theories, 
it is possible to discriminate between different modified 
gravity models from the growth of matter perturbations.

In order to quantify the difference between the two gravitational 
potentials $\Psi$ and $\Phi$, we define
\begin{equation}
\eta \equiv -\Phi/\Psi\,.
\label{etadef}
\end{equation}
On using the solutions (\ref{Psiso}) and (\ref{Phiso}), the anisotropy 
parameter reads
\begin{equation}
\eta \simeq \frac{(B_{8}D_{9}-A_{6}B_{7})(k/a)^{2}-B_{8}M^{2}}
{(B_{6}D_{9}-B_{7}^{2})(k/a)^{2}-B_{6}M^{2}}\,.
\label{etadef2}
\end{equation}
The effective gravitational potential associated with 
deviation of the light rays in CMB and weak lensing 
observations is given by \cite{lensing}
\begin{equation}
\Phi_{{\rm eff}}\equiv(\Psi-\Phi)/2\,,
\end{equation}
{}From Eqs.~(\ref{Psiso}), (\ref{Geff}), and (\ref{etadef}),
we obtain 
\begin{equation}
\Phi_{{\rm eff}} \simeq 
-4\pi G_{{\rm eff}}\frac{1+\eta}{2}\left(\frac{a}{k}\right)^{2}\rho_{m}\delta_m\,,
\label{Phieff}
\end{equation}
which is related to both $\delta_m$ and $\eta$.

\subsection{Growth of matter perturbations}

Introducing the matter density parameter 
$\Omega_m=\rho_m/(3M_{\rm pl}^2 H^2)$, 
we can write the matter perturbation equation 
(\ref{mattereqf}) in the form
\begin{equation}
\delta_{m}''+\left(2+\frac{H'}{H}\right)\delta_{m}'
-\frac{3}{2}\frac{G_{{\rm eff}}}{G}\Omega_{m}\delta_{m}\simeq0\,,
\label{delmeq}
\end{equation}
where a prime  represents a derivative with respect to $\ln a$.

Let us first consider a non-canonical scalar field 
described by the Lagrangian 
\begin{equation}
L=\frac{M_{\rm pl}^2}{2}R+P(\phi,X)\,,
\end{equation} 
in which case $G_2=P(\phi,X)$, $G_3=0$, 
$G_4=M_{\rm pl}^2/2$, and $G_5=0$. 
Since $B_6=B_8=2M_{\rm pl}^2$, 
$B_7=A_6=0$, and $D_9=2P_{X}$, it follows that 
$G_{\rm eff}=G$ and $\eta=1$ from 
Eqs.~(\ref{Geff}) and (\ref{etadef2}). 
During the matter-dominated epoch characterized by 
$\Omega_m=1$ and $H'/H=-3/2$, there is the 
growing-mode solution to Eq.~(\ref{delmeq}):
\begin{equation}
\delta_m \propto t^{2/3}\,.
\label{delmso}
\end{equation} 
In this regime, the effective gravitational potential 
(\ref{Phieff}) is constant. 
After the Universe enters the epoch of cosmic acceleration,
the growth rate of $\delta_m$ becomes smaller than 
that given in Eq.~(\ref{delmso}), so $\Phi_{\rm eff}$
starts to decay. Since $G_{\rm eff}$ is equivalent to 
$G$ for the models in the framework of GR, 
the difference of the growth rate between 
the models comes from the different background
expansion history. 
In the $\Lambda$CDM model characterized by $P=-\Lambda$, 
the growth rate $f \equiv \dot{\delta}_m/(H \delta_m)$ can be 
estimated as $f=(\Omega_m)^{\gamma}$ with $\gamma \simeq 0.55$ 
in the low-redshift regime ($z<1$) \cite{Stein}.
As long as the dark energy equation of state does not significantly 
deviate from $-1$, $\gamma$ is close to the value $0.55$ for 
the models in the framework of GR \cite{Linder,gamma2}.

As an example of modified gravity models, 
we consider BD theory described by the action (\ref{BDaction}). 
Since $B_6=2M_{\rm pl}\phi$, $B_7=2M_{\rm pl}$, 
$B_8=2M_{\rm pl}\phi$, $A_6=M_{\rm pl}$, and 
$D_9=-M_{\rm pl}\omega_{\rm BD}/\phi$, 
Eqs.~(\ref{Geff}) and (\ref{etadef2}) reduce to
\begin{eqnarray}
G_{{\rm eff}}&=&
\frac{\Mpl}{\phi}\frac{4+2\omega_{{\rm BD}}+2(\phi/\Mpl)(Ma/k)^{2}}
{3+2\omega_{{\rm BD}}+2(\phi/\Mpl)(Ma/k)^{2}}G\,,\\
\eta&=&
\frac{1+\omega_{{\rm BD}}+(\phi/\Mpl)(Ma/k)^{2}}
{2+\omega_{{\rm BD}}+(\phi/\Mpl)(Ma/k)^{2}}\,,\label{GeffBD}
\end{eqnarray}
where 
\begin{equation}
M^{2}=V_{\phi\phi}+\frac{\omega_{{\rm BD}}\Mpl}{\phi^{3}}
\left[\dot{\phi}^{2}-\phi\left(\ddot{\phi}+3H\dot{\phi}\right)\right]\,.
\end{equation}
In the $\omega_{\rm BD} \to \infty$ limit with $\phi \to M_{\rm pl}$,  
we obtain $G_{\rm eff} \to G$ and $\eta \to 1$, so the General 
Relativistic behavior can be recovered. 
The same property also holds for $M \to \infty$, as 
the scalar field does not propagate.

In the massless limit $M^2 \to 0$, it follows that 
$G_{{\rm eff}}\simeq(\Mpl/\phi)(4+2\omega_{{\rm BD}})G/(3+2\omega_{{\rm BD}})$
and $\eta\simeq(1+\omega_{{\rm BD}})/(2+\omega_{{\rm BD}})$, 
so the growth rates of $\delta_m$ and $\Phi_{\rm eff}$ are 
different from those in GR.
Since $\omega_{\rm BD}=0$ in metric $f(R)$ gravity, we have
$G_{{\rm eff}}\simeq(\Mpl/\phi)(4/3)G$ and 
$\eta \simeq 1/2$. The viable dark energy models based 
on $f(R)$ gravity \cite{fRviable1,fRviable2,fRviable3,fRviable4,fRviable5} 
are constructed in a way that the mass $M$ 
is large for $R \gg H_0^2$ and that $M$ decreases to 
the similar order to $H_0$ by today. 
There is a transition from the ``massive'' regime $M>k/a$ to 
the ``massless'' regime $M<k/a$, depending on the 
wavenumber $k$ \cite{fRviable3,fRviable4,GT09}.
If this transition happens in the deep matter era 
characterized by $H'/H \simeq -3/2$ and 
$\tilde{\Omega}_m=\rho_m/(3M_{\rm pl}\phi H^2) \simeq 1$, 
the growing-mode solution to Eq.~(\ref{delmeq}) during the 
``massless'' regime of metric $f(R)$ gravity is given by \cite{fRviable3}
\begin{equation}
\delta_m \propto t^{(\sqrt{33}-1)/6}\,,
\end{equation}
whose growth rate is larger than that in GR. 
This leaves an imprint for the measurement of red-shift space 
distortions in the galaxy power spectrum \cite{RSD}. 
{}From Eq.~(\ref{Phieff}), the effective gravitational coupling evolves as 
$\Phi_{\rm eff} \propto t^{(\sqrt{33}-5)/6}$.
This modification affects the weak lensing power spectrum 
as well as the ISW effect in CMB \cite{wlensing,Song}.

In other modified gravity models like covariant Galileons \cite{Kase}, 
the growth rate of perturbations is different from that in GR 
and $f(R)$ gravity.
Although the current observations are not enough to 
discriminate between different models precisely, 
we hope that future observations will allow us to do so.

\section{Conclusions}
\label{consec}

We have reviewed a framework for studying the most general 
four-dimensional gravitational theories with a single scalar 
degree of freedom.
The EFT of cosmological perturbations is useful for the unified
description of modified gravitational theories in that it can be
describe practically all single-field models proposed in the literature. 
This unified scheme can allow one to provide model-independent 
constraints on the properties of inflation/dark energy and to
put constraints on individual models consistent with observations.

Starting from the general action (\ref{action0}) that depends on 
the lapse $N$ and other three-dimensional scalar ADM variables, 
we have expanded the action up to quadratic order in cosmological 
perturbations about the FLRW background.
The choice of unitary gauge allows one to absorb dynamics of 
the field perturbation $\delta \phi$ into the gravitational sector.
Provided that the three conditions (\ref{elicon1})-(\ref{elicon3}) 
are satisfied, the second-order Lagrangian density reduces to the 
simple form (\ref{perlag}) with a single scalar degree of freedom 
characterized by the curvature perturbation $\zeta$. 
We have also shown that the quadratic action for tensor 
perturbations is given by Eq.~(\ref{tenlag}). 
In order to avoid ghosts and Laplacian instabilities of scalar 
and tensor perturbations, we require the conditions 
$Q_s>0$, $c_s^2>0$, $Q_t>0$, and $c_t^2>0$.

The most general scalar-tensor theories with second-order 
equations of motion--Horndeski theory--belong to a 
sub-class of the action (\ref{action0}) in the framework of EFT. 
The Horndeski Lagrangian can be expressed in terms of 
the ADM scalar quantities in the form (\ref{Ltotal}).
Using the relations (\ref{bac1})-(\ref{corres}) between the EFT 
variables appearing in the action (\ref{action2}) and the partial 
derivatives of the Lagrangian $L$ with respect to the ADM variables, 
we have shown that, up to quadratic order in perturbations, Horndeski 
theory corresponds to the action (\ref{action3}) with the additional 
condition $m_4^2=\mu_1^2$.
The dictionary between the EFT variables and the functions $G_{i}(\phi,X)$
in Horndeski theory is given by Eqs.~(\ref{effq1})-(\ref{corre9}).

In Sec.~\ref{infsec} we have also derived the power spectra of scalar 
and tensor perturbations generated during inflation for
general second-order theories satisfying the conditions 
(\ref{elicon1})-(\ref{elicon3}). The formulas (\ref{scalarpower}) 
and (\ref{tensorpower}) 
cover a wide variety of modified gravitational theories presented 
in Sec.~\ref{conmodel}, so they can be used for constraining 
each inflationary model from the CMB observations (along the lines of 
Ref.~\cite{Kuroyanagi}). In particular, it will be of interest to 
discriminate between a host of single-field inflationary models 
from the precise B-mode polarization data available in the future.

In Sec.~\ref{darksec} we have applied the EFT of cosmological 
perturbations to dark energy in the presence of a barotropic perfect fluid.
The background cosmology is described by three time-dependent 
functions $f$, $\Lambda$, and $c$, with which different 
models can be distinguished from the evolution of the dark 
energy equation of state.
In Horndeski theory, we have obtained the effective gravitational coupling 
(\ref{Geff}) appearing in the matter perturbation equation (\ref{mattereqf})
under the quasi-static approximation on sub-horizon scales.
Together with the effective gravitational potential given in Eq.~(\ref{Phieff}), 
it will be possible to discriminate between different modified gravity models from 
the observations of large-scale structures, weak lensing, and CMB.

While we have studied the effective single-field scenario in unitary gauge, 
another scalar degree of freedom can be also taken into account 
in the action (\ref{action0}) \cite{Gergely}. 
Such a second scalar field can be potentially responsible 
for dark matter. It will be of interest to provide a unified framework 
for understanding the origins of inflation, dark energy, and dark matter.

\section*{Acknowledgments}

The author is grateful to the organizers of the 7th Aegean 
Summer School for wonderful hospitality.
The author thanks Antonio De Felice, Laszlo Arpad Gergely, 
and Federico Piazza for useful discussions.
This work was supported by Grant-in-Aid for Scientific Research 
Fund of the JSPS (No.\, 30318802) and 
Grant-in-Aid for Scientific Research on Innovative 
Areas (No.\, 21111006).


\end{document}